\documentclass{article}

\usepackage{arxiv}

\usepackage[utf8]{inputenc} 
\usepackage[T1]{fontenc}    
\usepackage{hyperref}       
\usepackage{url}            
\usepackage{booktabs}       
\usepackage{amsfonts}       
\usepackage{nicefrac}       
\usepackage{microtype}      
\usepackage{lipsum}
\usepackage{graphicx}
\graphicspath{ {./images/} }

\newcommand\setItemnumber[1]{\setcounter{enumi}{\numexpr#1-1\relax}}

\title{A methodology for the measurment of track geometry based on computer vision and inertial sensors}

\author{
 José L. Escalona \\
  Dept. of Mechanical and Manufacturing Engineering\\
  University of Seville\\
  \texttt{escalona@us.es} \\
}

\begin{document}
\maketitle
\begin{abstract}
This document describes the theory used for the calculation of track geometric irregularities on a \textit{Track Geometry Measuring System} (TGMS) to be installed in railway vehicles. The TGMS includes a computer for data acquisition and process, a set of sensors including an \textit{inertial measuring unit} (IMU, 3D gyroscope and 3D accelerometer), two video cameras and an encoder.
\end{abstract}

\keywords{Computer Vision \and Motion Tracking \and Multibody System Dynamics \and Laser Projector \and Zhang Calibration Method}

\section{Introduction}

 There is many commercial equipment used for track geometry measurement that are based on inertial and optical sensors. However, to the author best knowledge, the detailed methods used for the calculation of track irregularities remain unpublished. The main features of the proposed system are:

\begin{enumerate}
	\item It is capable to measure track alignment, vertical profile, cross-level, gauge, twist and rail-head profile using non-contact technology.
	\item It can be installed in line railway vehicles. It is compact and low cost. Provided that the equipment sees the rail heads when the vehicle is moving, it can be installed in any body of the vehicle: at the wheelsets level, above primary suspension (bogie frame) or above the secondary suspension (car body).
\end{enumerate}

This document includes the following sections:

\begin{enumerate}
	\setItemnumber{2} 
	\item Description of the TGMS
	\item Kinematics of the irregular track and the railway vehicle
	\item Kinematics of the computer vision
	\item Detecting the rail cross-section in a camera frame
	\item Equations for geometry measurement
	\item Measurement of TGMS to TF relative motion
	\item Odometry algorithm
	\item Sensor fusion algorithm to find TGMS to TF relative angles
	\item Calibration of the cameras
	\item Summary of the measurement of track irregularities
	\item Final considerations
\end{enumerate}

\section{Description of the TGMS} \label{sec:TGMSdescription}

\begin{figure}[htbp!]
	\centering
	\includegraphics[width=0.7\linewidth]{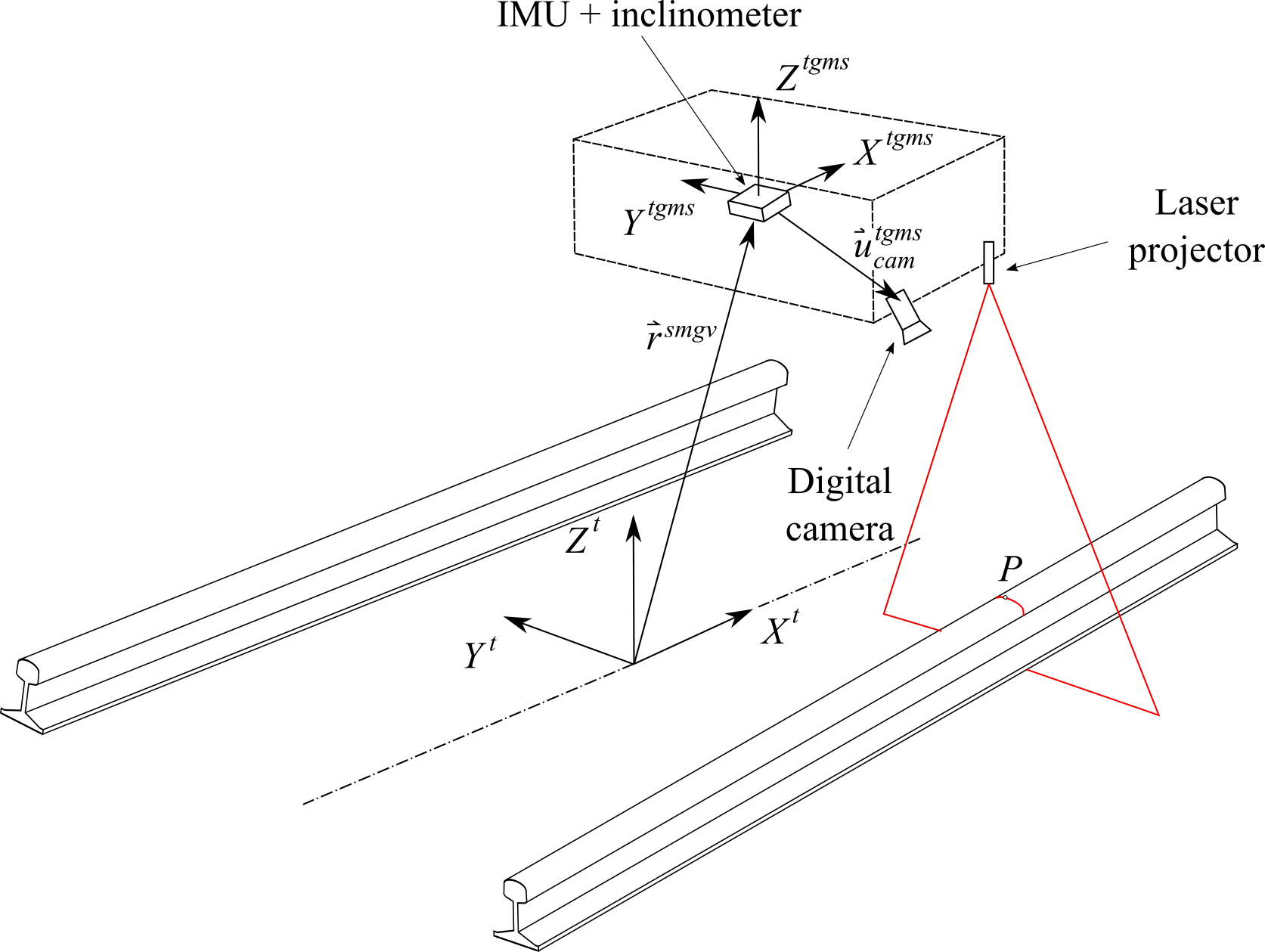}
	\caption{Kinematics of the TGMS installed in a vehicle moving along the track}
	\label{fig:fig2_1}
\end{figure}

The TGMS sketched in Figs. \ref{fig:fig2_1} (only right-side equipment in shown here) and \ref{fig:fig2_2} comprises:

\begin{enumerate}
	\item Two video cameras
	\item Two laser line-projectors
	\item An IMU
	\item A signal to detect the position along the track
\end{enumerate}

Regarding point 4, this signal: (1) may come from the vehicle odometer, if any, or (2) it can be obtained using a GNSS sensor, or (3) may be obtained using an encoder installed in a wheel of the vehicle. Optionally, the TGMS may include also:

\begin{enumerate}
	\setItemnumber{6} 
	\item A two axis inclinometer
\end{enumerate}

It is important that the cameras, lasers and IMU are installed in a solid that can be considered as a rigid body when moving with the vehicle. The lasers and cameras must be equipped with orientation mechanisms that can be fully locked when the TGMS is working. The laser projectors draw red lines (when using a red laser) in the rail-heads (one on the left, one on the right) that are filmed by the video cameras. The information provided by the position and orientation of the read lines in the frames, together with the acceleration and angular velocity acquired with the IMU, are used to find the track geometry irregularities.

\textbf{\begin{figure}[htbp!]
	\centering
	\includegraphics[width=0.8\linewidth]{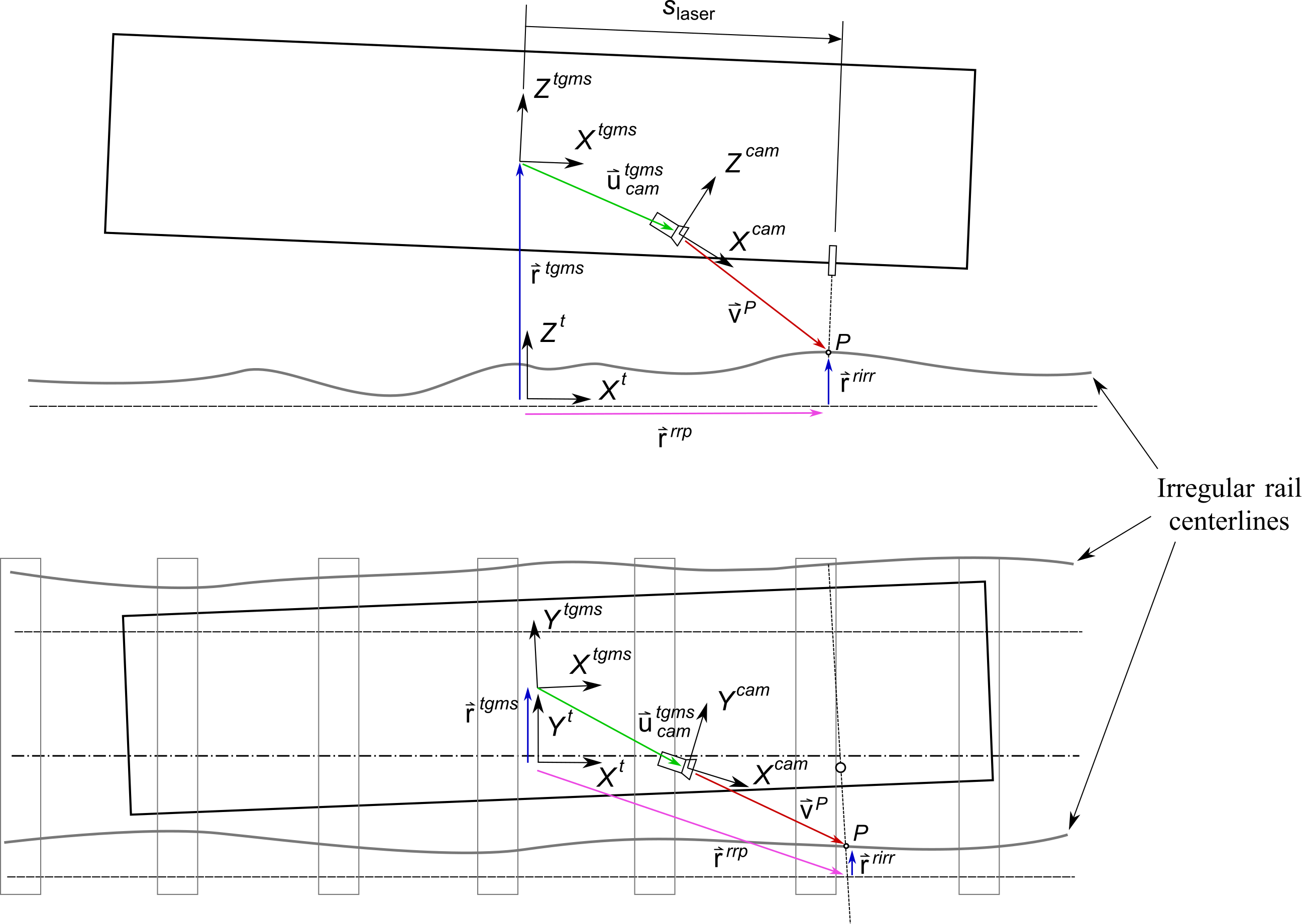}
	\caption{Side and top views of the TGMS}
	\label{fig:fig2_2}
\end{figure}}

\section{Kinematics of the irregular track and the railway vehicle} \label{sec:kinematics}

This section includes the kinematic description of the rail geometry as a combination of an \textit{ideal geometry} and the \textit{irregularities}, and the kinematic description of an arbitrary body moving along the track, like the TGMS.  Before presenting the kinematics, the different frames that are used and the nomenclature used to describe vectors, matrices and their components are described.

\subsection{Nomenclature} \label{sec:nomenclature}

\begin{figure}[htbp!]
	\centering
	\includegraphics[width=0.6\linewidth]{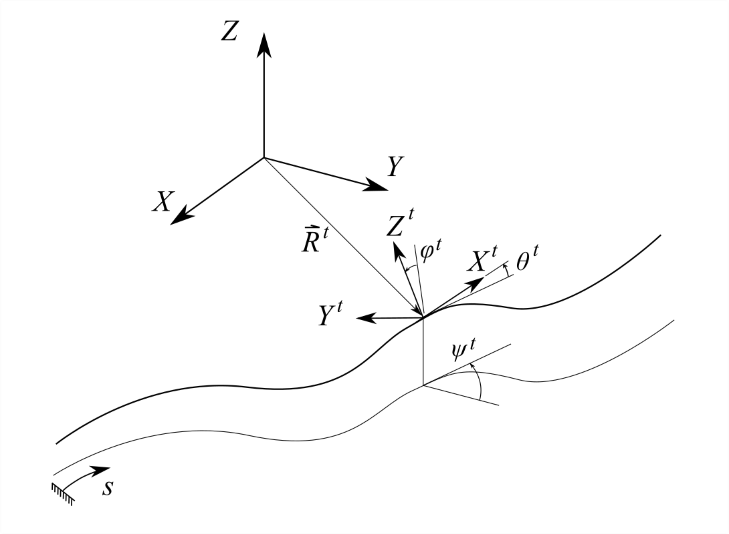}
	\caption{Ideal track centerline}
	\label{fig:fig3_1}
\end{figure}

\begin{figure}[htbp!]
	\centering
	\includegraphics[width=0.6\linewidth]{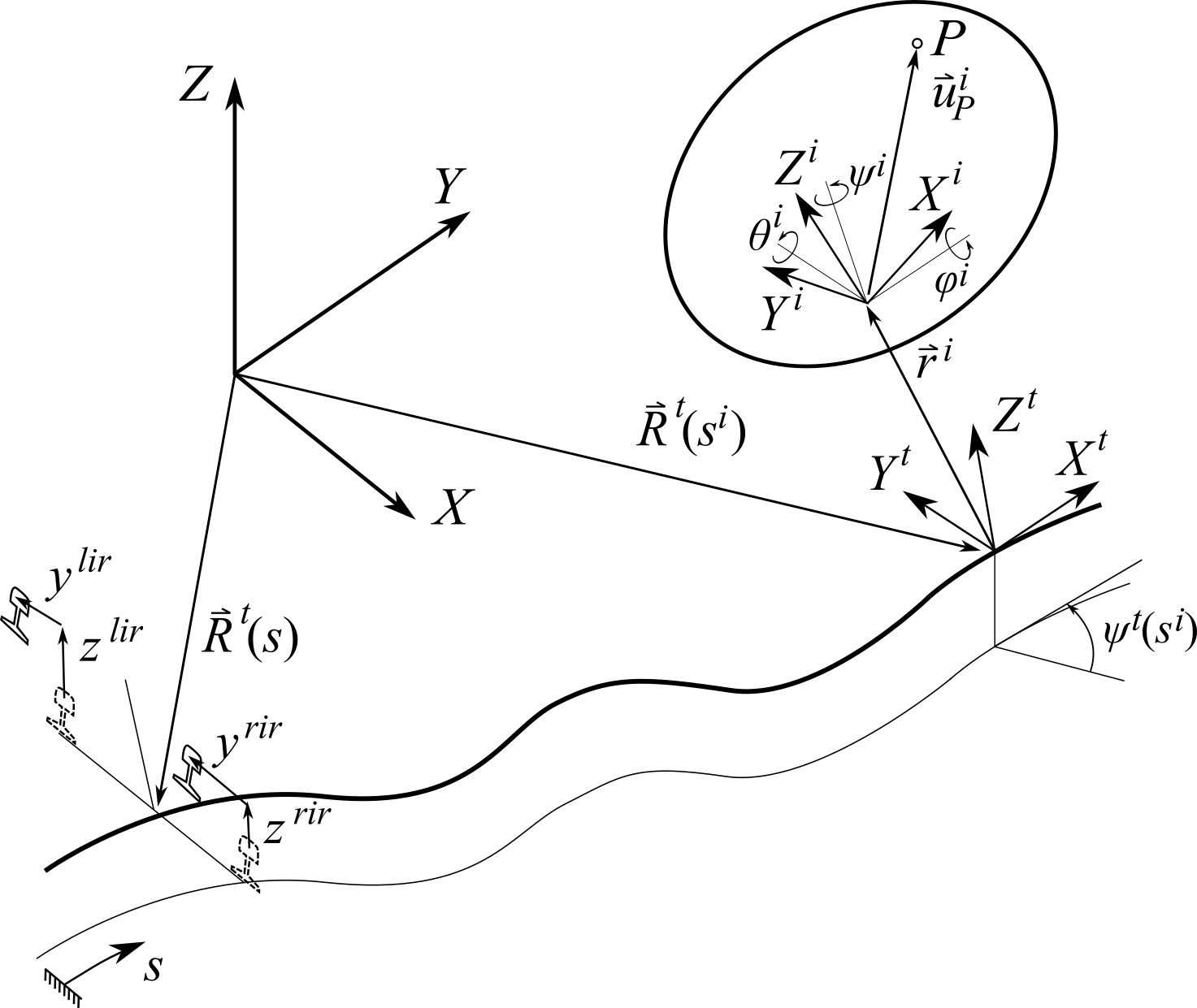}
	\caption{Kinematics of irregular track and moving body}
	\label{fig:fig3_2}
\end{figure}

\begin{figure}[htbp!]
	\centering
	\includegraphics[width=0.7\linewidth]{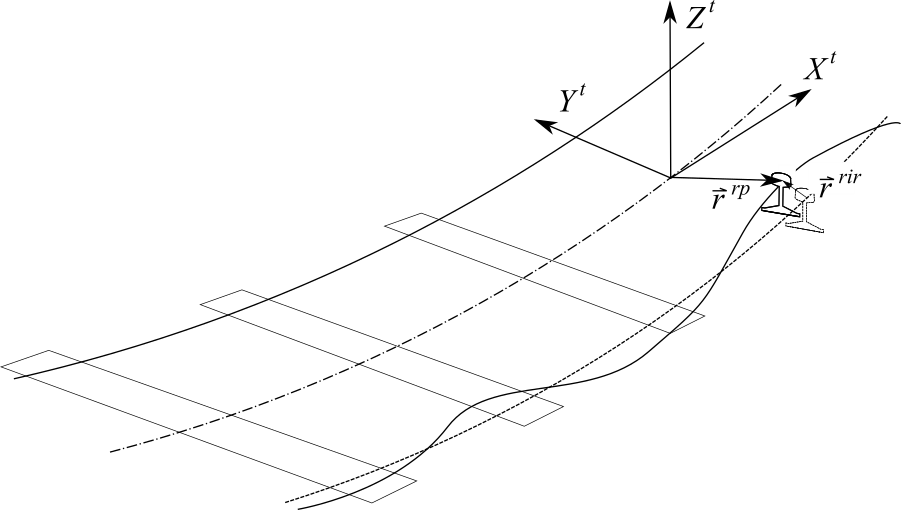}
	\caption{Rail centerline irregularity}
	\label{fig:fig3_3}
\end{figure}

\begin{figure}[htbp!]
	\centering
	\includegraphics[width=0.6\linewidth]{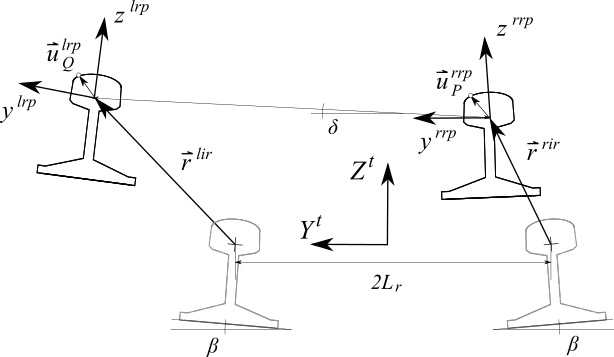}
	\caption{Kinematics of the track cross-section}
	\label{fig:fig3_4}
\end{figure}

As shown in Figs. \ref{fig:fig3_1} - \ref{fig:fig3_4}, four different frames are used in railroad kinematics:

\begin{enumerate}
	\item The inertial and \textit{global frame} (GF) $<X, Y, Z>$. It is a frame fixed in space.
	\item The \textit{track frame} (TF) $<X^t, Y^t, Z^t>$. It is not a single frame but a field defined for each value of the arc-length coordinate along the track $s$. The position ${{\bf{R}}^t}\left( s \right)$ and orientation matrix ${{\bf{A}}^t}\left( s \right)$ of the TF with respect to the GF are functions of an arc-length coordinate s along the center line of the ideal track (without irregularities). These functions are implemented computationally in the \textit{Track Preprocessor}.
	\item The body frame (BF) $<X^i, Y^i, Z^i>$ of each body $i$. It is a frame rigidly attached to the body. In this document, the body i is the TGMS. The body frame of the TGMS is denoted as $<X^{tgms}, Y^{tgms}, Z^{tgms}>$
	\item The rail profile frames. \textit{Left rail-profile frame} (LRP), $<X^{lrp}, Y^{lrp}, Z^{lrp}>$, and \textit{right-profile frame} (RRP), $<X^{rrp}, Y^{rrp}, Z^{rrp}>$. These frames are not a unique frames but fields defined for each value of the arc-length coordinate along the track $s$. These frames are rigidly attached to the rail-heads.
\end{enumerate}

The definition of the TF is such that the $X^t$ axis is tangent to the track ideal centerline, the $Y^t$ axis is perpendicular to $X^t$ and connects the origin $O^{lrp}$ of the LRP and the origin $O^{rrp}$ of the RRP in the ideal track geometry (with no track irregularities) and the $Z^t$ axis is perpendicular to both $X^t$ and $Y^t$. Therefore, the TF is not the Frenet frame of the ideal track centerline. Each body $i$ moving along the track has an associated TF at each instant of time. Its position and orientation can be obtained substituting the position of the body along the track, $s^i(t)$, in the functions ${{\bf{R}}^t}\left( s \right)$ and ${{\bf{A}}^t}\left( s \right)$.
In this document, vector values are defined with a symbol in italic with an arrow. Position vectors are defined using the following nomenclature:

\begin{enumerate}
	\item $\vec R$ is a position vector with respect to the GF.
    \item $\vec r$ is a position vector with respect to the TF, with the exception of the irregularity vectors $\vec r^{lir}$ and $\vec r^{rir}$, shown in Figs. \ref{fig:fig3_3} and \ref{fig:fig3_4}, that has the origin at the LRP and RRP of the ideal position of the rail heads.
    \item $\vec u$ is a position vector with respect to the BF, LRP or RRP.
\end{enumerate}

The column matrix that contains the components of a vector in a frame is defined using bold symbols, in general with a “hat”, as follows:

\begin{enumerate}
	\item Bold symbols without hat, like $\bf{v}$, means the $3\times1$ column matrix that contains the components of vector $\vec v$ in the GF.
	\item Bold symbols with ‘bar’ superscript, like ${\bf{\bar v}}$, means the $3\times1$ column matrix that contains the components of vector $\vec v$ in the TF.
	\item Bold symbols with ‘arc’ superscript, like ${\bf{\hat v}}$, means the $3\times1$ column matrix that contains the components of vector $\vec v$ in the BF, LRP or RRP.
\end{enumerate}

Symbols representing vectors may include superscripts and subscripts, as vector $\vec u_P^i$ in Fig. \ref{fig:fig3_2}. In that case, the subscript means the name of the point and the superscript means the body to which the point belongs. Following the nomenclature defined above, ${\bf{u}}_P^i$, ${\bf{\bar u}}_P^i$ and ${\bf{\hat u}}_P^i$, mean the column matrices ($3\times1$) of the vector $\vec u_P^i$ in the GF, TF and BF, respectively.

Rotation matrices are expressed with symbol $\bf{A}$ and two superscripts separated by comma. For example, ${{\bf{A}}^{t,i}}$ is the $3\times3$ rotation matrix from the BF of body $i$ to the TF (whose symbol is $t$). It is easy to follow that ${\bf{\bar u}}_P^i = {{\bf{A}}^{t,i}}{\bf{\hat u}}_P^i$. Rotation matrices with just one superscript are rotation matrices with respect to the GF. For example, ${{\bf{A}}^{t}}$ and ${{\bf{A}}^{i}}$ are the rotation matrices of the TF and the BF with respect to the GF, respectively. It is easy to follow that: ${\bf{u}}_P^i = {{\bf{A}}^i}{\bf{\hat u}}_P^i$, ${\bf{u}}_P^i = {{\bf{A}}^i}{\bf{\hat u}}_P^i$ and ${{\bf{A}}^i} = {{\bf{A}}^t}{{\bf{A}}^{t,i}}$.

\subsection{Kinematics of the ideal track centerline} \label{sec:centerline}

Track geometry is the superposition of the ideal geometry and the irregularities. The components of the absolute position vector of an arbitrary point on the ideal track centerline with respect to an inertial and global frame is a function of the arc-length $s$, as follows: 

\begin{equation} \label{Eq3_1}
{{\bf{R}}^t}\left( s \right) = \left[ {\begin{array}{*{20}{c}}
{R_x^t\left( s \right)}\\
{R_y^t\left( s \right)}\\
{R_z^t\left( s \right)}
\end{array}} \right]
\end{equation}

where ${{\bf{R}}^t}$ contains the components of vector ${\vec R^t}$ shown in Fig. \ref{fig:fig3_1}. The geometry of the track centerline 3D-curve is defined by the \textit{horizontal profile} and the $vertical profile$. Both profiles are defined using sections of variable length. Points between two sections are called vertices. Horizontal profile vertices do not necessary coincide with vertical profile vertices. Horizontal profile includes three types of sections: tangent (straight), curve (circular) and transitions (clothoid). Vertical profile includes two types of sections:  constant-slope (straight) and transitions (cubic).

At each track section, the track centerline geometry is characterized by the following geometric values:

\begin{tabular}{ l l }
    Horizontal curvature: & ${\rho _h}$ \\
    Vertical curvature: & ${\rho _v}$ \\
    Twist curvature: & ${\rho _{tw}}$ \\
    Spatial-derivative of horizontal curvature: & ${\rho _h}^\prime $ \\
    Vertical slope: & ${\alpha _v}$
\end{tabular}

Horizontal and vertical profile sections show the following values of these parameters:

\textbf{Horizontal profile:}

\begin{tabular}{ l l l l}
    Straight section: & ${\rho _h}=0$, & ${\rho _{tw}}=0$, & ${\rho _h}^\prime =0$, \\
    Circular section: & ${\rho _h} = {1 \mathord{\left/
 {\vphantom {1 {{R_h}}}} \right.
 \kern-\nulldelimiterspace} {{R_h}}}$, & ${\rho _{tw}}=0$, & ${\rho _h}^\prime =0$, \\
    Transition section: & ${\rho _h} = \left( {{1 \mathord{\left/
 {\vphantom {1 {{R_{h1}}}}} \right.
 \kern-\nulldelimiterspace} {{R_{h1}}}}} \right) + {f_{lin}}\left( s \right)\left( {{1 \mathord{\left/
 {\vphantom {1 {{R_{h2}} - {1 \mathord{\left/
 {\vphantom {1 {{R_{h1}}}}} \right.
 \kern-\nulldelimiterspace} {{R_{h1}}}}}}} \right.
 \kern-\nulldelimiterspace} {{R_{h2}} - {1 \mathord{\left/
 {\vphantom {1 {{R_{h1}}}}} \right.
 \kern-\nulldelimiterspace} {{R_{h1}}}}}}} \right),$, & ${\rho _{tw}} = {{\left( {{\varphi _{p2}} - {\varphi _{p1}}} \right)} \mathord{\left/
 {\vphantom {{\left( {{\varphi _{p2}} - {\varphi _{p1}}} \right)} {{L_{ht}}}}} \right.
 \kern-\nulldelimiterspace} {{L_{ht}}}},$, & ${\rho _h}^\prime  = {{\left( {{1 \mathord{\left/
 {\vphantom {1 {{R_{h2}} - {1 \mathord{\left/
 {\vphantom {1 {{R_{h1}}}}} \right.
 \kern-\nulldelimiterspace} {{R_{h1}}}}}}} \right.
 \kern-\nulldelimiterspace} {{R_{h2}} - {1 \mathord{\left/
 {\vphantom {1 {{R_{h1}}}}} \right.
 \kern-\nulldelimiterspace} {{R_{h1}}}}}}} \right)} \mathord{\left/
 {\vphantom {{\left( {{1 \mathord{\left/
 {\vphantom {1 {{R_{h2}} - {1 \mathord{\left/
 {\vphantom {1 {{R_{h1}}}}} \right.
 \kern-\nulldelimiterspace} {{R_{h1}}}}}}} \right.
 \kern-\nulldelimiterspace} {{R_{h2}} - {1 \mathord{\left/
 {\vphantom {1 {{R_{h1}}}}} \right.
 \kern-\nulldelimiterspace} {{R_{h1}}}}}}} \right)} {{L_{ht}}}}} \right.
 \kern-\nulldelimiterspace} {{L_{ht}}}}$, \\
\end{tabular}

where $R_h$ is the curve radius, ${f_{lin}}\left( s \right)$ is a linear function of the arc-length that is zero at the at the straight end and one at the curved end, ${\varphi _p}$ is the cant angle at the curved section and ${L_{ht}}$ is the length of the transition section. Subscripts ‘1’ and ‘2’ used in the definition of the curvatures of the transition section are related to the anterior and posterior segments along the track, respectively.

\textbf{Vertical profile:}

\begin{tabular}{ l l l l}
    Straight section: & ${\alpha _v} = $ constant, & ${\rho _v}=0$, \\
    Transition section: & ${\alpha _v} = {\alpha _{v1}} + {f_{lin}}\left( s \right)\left( {{\alpha _{v2}} - {\alpha _{v1}}} \right)$, & ${\rho _v} = {{\left( {{\alpha _{v2}} - {\alpha _{v1}}} \right)} \mathord{\left/
 {\vphantom {{\left( {{\alpha _{v2}} - {\alpha _{v1}}} \right)} {{L_{vt}}}}} \right.
 \kern-\nulldelimiterspace} {{L_{vt}}}}$, \\
\end{tabular}

where ${\alpha _{v2}}$ and ${\alpha _{v1}}$ are the slopes of the straight section before and after the transition, and ${L_{vt}}$  is the length of the transition section.

The list of sections of the horizontal and vertical profiles of a given track, including the value of the geometric parameters given above, allows the calculation of the function ${{\bf{R}}^t}\left( s \right)$ given in Eq. \ref{fig:fig3_1}. This function use to be implemented in a track-preprocessor that is a very important part of the railroad dynamic simulation codes. The orientation of the track centerline can also be obtained as a function of $s$ as explained next.  

Figure  \ref{fig:fig3_1} shows the TF $<X^t, Y^t, Z^t>$ associated with the track centerline at each value of $s$. The orientation of the TF with respect to a GF can be measured with the Euler angles ${\psi ^{t}}$  ($azimut$ or $heading$ angle), ${\theta ^{\,t}}$  (vertical slope, positive when downwards in the forward direction) and ${\varphi ^{\,t}}$  ($cant$ or $superelevation$ $angle$). The rotation matrix from the TF to the GF is given by:

\begin{equation} \label{Eq3_2}
{{\bf{A}}^t}\left( s \right) = \left[ {\begin{array}{*{20}{c}}
{{\mathop{\rm c}\nolimits} {\theta ^{\,t}}{\rm{c}}{\psi ^t}}&{{\rm{s}}{\varphi ^t}{\rm{s}}{\theta ^{\,t}}{\mathop{\rm c}\nolimits} {\psi ^t} - {\rm{c}}{\varphi ^t}{\rm{s}}{\psi ^t}}&{{\rm{s}}{\varphi ^t}{\rm{s}}{\psi ^t} + {\mathop{\rm c}\nolimits} {\varphi ^t}{\rm{s}}{\theta ^{\,t}}{\mathop{\rm c}\nolimits} {\psi ^t}}\\
{{\mathop{\rm c}\nolimits} {\theta ^{\,t}}{\rm{s}}{\psi ^t}}&{{\mathop{\rm c}\nolimits} {\varphi ^t}{\mathop{\rm c}\nolimits} {\psi ^t} + {\rm{s}}{\varphi ^t}{\rm{s}}{\theta ^{\,t}}{\rm{s}}{\psi ^t}}&{{\mathop{\rm c}\nolimits} {\varphi ^t}{\rm{s}}{\theta ^{\,t}}{\rm{s}}{\psi ^t} - {\rm{s}}{\varphi ^t}{\rm{c}}{\psi ^t}}\\
{ - {\rm{s}}{\theta ^{\,t}}}&{{\rm{s}}{\varphi ^t}{\rm{c}}{\theta ^{\,t}}}&{{\rm{c}}{\varphi ^t}{\rm{c}}{\theta ^{\,t}}}
\end{array}} \right]
\end{equation}

The azimut ${\psi ^{t}}$ can have any arbitrary value, however, the slope ${\theta ^{\,t}}$ and cant ${\varphi ^{\,t}}$ angles can be considered as small angles, such that the rotation matrix from the TF to the GF can be approximated to: 

\begin{equation} \label{Eq3_3}
{{\bf{A}}^t}\left( s \right) \simeq \left[ {\begin{array}{*{20}{c}}
{{\rm{c}}{\psi ^t}}&{ - {\rm{s}}{\psi ^t}}&{{\varphi ^t}{\rm{s}}{\psi ^t} + {\theta ^t}{\mathop{\rm c}\nolimits} {\psi ^t}}\\
{{\rm{s}}{\psi ^t}}&{{\mathop{\rm c}\nolimits} {\psi ^t}}&{{\theta ^t}{\rm{s}}{\psi ^t} - {\varphi ^t}{\rm{c}}{\psi ^t}}\\
{ - {\theta ^t}}&{{\varphi ^t}}&1
\end{array}} \right]
\end{equation}

An ideal body that moves along the track taking the same orientation as the track frame with a forward velocity   $V$ and a forward acceleration $\dot V$ has the following absolute velocity and acceleration: 

\begin{equation} \label{Eq3_4}
\bar{\bf{\dot{R}}}^t = \left[ {\begin{array}{*{20}{c}}
V\\
0\\
0
\end{array}} \right],\,\,\,\,\,\bar{\bf{\ddot{R}}}^t = \left[ {\begin{array}{*{20}{c}}
{\dot V}\\
{{\rho _h}{V^2}}\\
{ - {\rho _v}{V^2}}
\end{array}} \right]
\end{equation}

where these arrays contain the first and second time-derivatives of vector ${\vec R^t}$ in the track frame. Similarly, the absolute angular velocity and the absolute angular acceleration of that body are given by: 

\begin{equation} \label{Eq3_5}
{{\bf{\bar \omega }}^t} = \left[ {\begin{array}{*{20}{c}}
{{\rho _{tw}}V}\\
{{\rho _v}V}\\
{{\rho _h}V}
\end{array}} \right],\,\,\,\,\,{{\bf{\bar \alpha }}^t} = \left[ {\begin{array}{*{20}{c}}
{{\rho _{tw}}\dot V}\\
{{\rho _v}\dot V}\\
{{\rho _h}\dot V + {{\rho '}_h}{V^2}}
\end{array}} \right]
\end{equation}

where these arrays contain the components of the angular velocity and acceleration vectors also in the TF.

\subsection{Kinematics of the irregular track} \label{sec:irregularTrack}

Figure \ref{fig:fig3_3} shows the relative position of the irregular right rail centerline with respect to the TF defined in the previous subsection. Figure \ref{fig:fig3_4} shows the displacement of the rail heads due to irregularity in a cross-section of the track (${Y^t} - {Z^t}$ plane). The irregularity vectors ${\vec r^{lir}}$ ($lir$, $l$eft rail $ir$regularity) and ${\vec r^{rir}}$ ($rir$, $r$ight rail $ir$regularity) describe the displacement of the rail centerlines with respect to their ideal positions. The components of these vectors in the TF are functions of $s$, given by:

\begin{equation} \label{Eq3_6}
{{\bf{\bar r}}^{lir}} = \left[ {\begin{array}{*{20}{c}}
0\\
{{y^{lir}}}\\
{{z^{lir}}}
\end{array}} \right],\,\,\,\,\,{{\bf{\bar r}}^{rir}} = \left[ {\begin{array}{*{20}{c}}
0\\
{{y^{rir}}}\\
{{z^{rir}}}
\end{array}} \right]
\end{equation}

In the railway industry, the following four combinations of the rail head centerlines irregularities are measured:

\begin{tabular}{ l l }
    Alignment ($al$): & $al = \frac{{{y^{lir}} + {y^{rir}}}}{2}$ \\
    Vertical profile ($vp$): & $vp = \frac{{{z^{lir}} + {z^{rir}}}}{2}$ \\
    Gauge variation ($gv$): & $gv = {y^{lir}} - {y^{rir}}$ \\
    Cross level ($cl$): & $cl = {z^{lir}} - {z^{rir}}$ \\
\end{tabular}

The orientation of the rail head frames with respect to the TF is given by the following rotation matrices:

\begin{equation} \label{Eq3_7}
\begin{array}{l}
{{\bf{A}}^{t,lrp}} = \left[ {\begin{array}{*{20}{c}}
1&0&0\\
0&{\cos \left( {\beta  + \delta } \right)}&{ - \sin \left( {\beta  + \delta } \right)}\\
0&{\sin \left( {\beta  + \delta } \right)}&{\cos \left( {\beta  + \delta } \right)}
\end{array}} \right],\\
{{\bf{A}}^{t,rrp}} = \left[ {\begin{array}{*{20}{c}}
1&0&0\\
0&{\cos \left( { - \beta  + \delta } \right)}&{ - \sin \left( { - \beta  + \delta } \right)}\\
0&{\sin \left( { - \beta  + \delta } \right)}&{\cos \left( { - \beta  + \delta } \right)}
\end{array}} \right],
\end{array}
\end{equation}

where $\beta $ is the orientation angle of the rail profiles and $\delta  = {{\left( {{z^{lir}} - {z^{rir}}} \right)} \mathord{\left/
 {\vphantom {{\left( {{z^{lir}} - {z^{rir}}} \right)} {2{L_r}}}} \right.
 \kern-\nulldelimiterspace} {2{L_r}}}$ is the linearized rotation angle due to the irregularity. Both angles can be observed in Fig. \ref{fig:fig3_4}.

The absolute position vectors of two points, $P$ and $Q$, defined in the right and left rail heads, respectively, are given by:

\begin{equation} \label{Eq3_8}
\begin{array}{l}
\vec R_P^{rrp} = {{\vec R}^t} + {{\vec r}^{rrp}} + {{\vec r}^{rir}} + \vec u_P^{rrp}\\
\vec R_Q^{lrp} = {{\vec R}^t} + {{\vec r}^{lrp}} + {{\vec r}^{lir}} + \vec u_Q^{lrp}
\end{array}
\end{equation}

\begin{figure}[htbp!]
	\centering
	\includegraphics[width=0.7\linewidth]{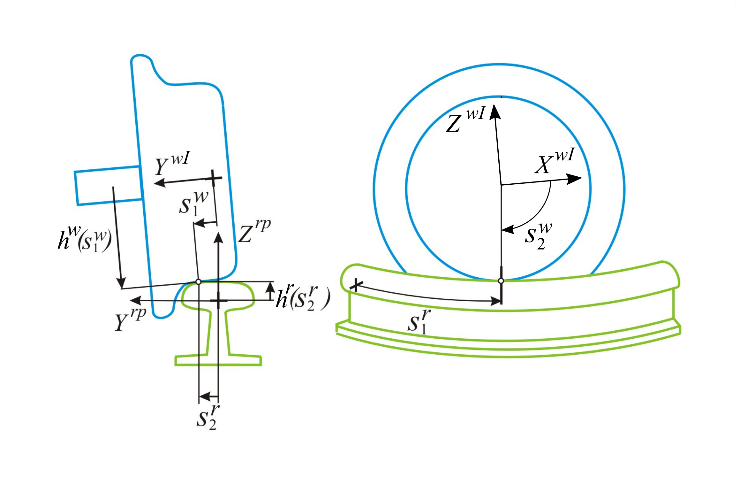}
	\caption{Wheel profile and rail profile geometry}
	\label{fig:fig3_5}
\end{figure}

The components of these vectors in the global frame are given by:

\begin{equation} \label{Eq3_9}
\begin{array}{l}
{\bf{R}}_P^{rrp} = {{\bf{R}}^t} + {{\bf{A}}^t}\left( {{{{\bf{\bar r}}}^{rrp}} + {{{\bf{\bar r}}}^{rir}} + {{\bf{A}}^{t,rrp}}{\bf{\hat u}}_P^{rrp}} \right)\\
{\bf{R}}_Q^{lrp} = {{\bf{R}}^t} + {{\bf{A}}^t}\left( {{{{\bf{\bar r}}}^{lrp}} + {{{\bf{\bar r}}}^{lir}} + {{\bf{A}}^{t,lrp}}{\bf{\hat u}}_Q^{lrp}} \right)
\end{array}
\end{equation}

where  ${\bf{\hat u}}_P^{rrp}$ and ${\bf{\hat u}}_Q^{lrp}$ contain the components of the position vector of points $P$ and $Q$ in the rail head profiles as shown in Fig. \ref{fig:fig3_4}. These vectors are parametrized following the rail head profile geometry:

\begin{equation} \label{Eq3_10}
{\bf{\hat u}}_P^{rrp}{\rm{ = }}\left[ {\begin{array}{*{20}{c}}
0\\
{s_2^{rr}}\\
{{h^r}\left( {s_2^{rr}} \right)}
\end{array}} \right],\,\,\,\,{\bf{\hat u}}_Q^{lrp}{\rm{ = }}\left[ {\begin{array}{*{20}{c}}
0\\
{s_2^{lr}}\\
{{h^r}\left( {s_2^{lr}} \right)}
\end{array}} \right]
\end{equation}

where $lr$ and $rr$ stand for "left rail" and "right rail", and $h^r$ is the function that defines the rail head profile, as shown in Fig. \ref{fig:fig3_5}.

\subsection{Kinematics of a body moving along the track} \label{sec:bodyAlongTrack}
The coordinates used to describe the position and orientation of an arbitrary body $i$, as shown in Fig. \ref{fig:fig3_2}, or the TGMS shown in Figs. \ref{fig:fig2_1} and \ref{fig:fig2_2}, moving along the track are:

\begin{equation} \label{Eq3_11}
{{\bf{q}}^i} = {\left[ {\begin{array}{*{20}{c}}
{{s^i}}&{r_y^i}&{r_z^i}&{{\varphi ^i}}&{{\theta ^i}}&{{\psi ^i}}
\end{array}} \right]^T}
\end{equation}

where $s^i$ is the arc-length along the track of the position of the body, $r_y^i$ and $r_z^i$ are the non-zero components of the position vector $\vec r^i$ of the BF with respect to the TF, this is ${{\bf{\bar r}}^i} = {\left[ {\begin{array}{*{20}{c}}
0&{r_y^i}&{r_z^i}
\end{array}} \right]^T}$  , and ${\varphi ^i},\,\,{\theta ^i}\,\,{\rm{and}}\,\,{\psi ^i}$ are three Euler angles (roll, pitch and yaw, respectively) that define the orientation of the BF with respect to the TF. These angles are assumed to be small, such that the following kinematic linearization is used:

\begin{equation} \label{Eq3_12}
{{\bf{A}}^{t,i}} \simeq \left[ {\begin{array}{*{20}{c}}
1&{ - {\psi ^i}}&{{\theta ^i}}\\
{{\psi ^i}}&1&{ - {\varphi ^i}}\\
{ - {\theta ^i}}&{{\varphi ^i}}&1
\end{array}} \right]
\end{equation}

The absolute position vector of point $P$ that belongs to body $i$, as shown in Fig. \ref{fig:fig3_2}, is given by:

\begin{equation} \label{Eq3_13}
\vec R_P^i = {\vec R^t} + {\vec r^i} + {\vec u_P^i}
\end{equation}

The absolute velocity and acceleration of point $P$ are given by:

\begin{equation} \label{Eq3_14}
\dot{\vec{R}}_P^i = \dot{\vec R^t} + {\vec r'^i} + {\vec \omega ^i} \times {\vec r^i} + {\vec \omega ^j} \times \vec u_P^j
\end{equation}

\begin{equation} \label{Eq3_15}
\ddot{\vec R^i}_P = \ddot {\vec R^t} + {\vec r''^i} + {\vec \alpha ^t} \times {\vec r^i} + {\vec \omega ^t} \times \left( {{{\vec \omega }^t} \times {{\vec r}^i}} \right) + 2{\vec \omega ^t} \times {\vec r'^i} + {\vec \alpha ^i} \times \vec u_P^i + {\vec \omega ^i} \times \left( {{{\vec \omega }^i} \times \vec u_P^i} \right)
\end{equation}

where symbol “prima” next to a vector, like ${\vec r'^i}$, means in this context (it is commonly used in mechanics for space-derivative) the time-derivative of the vector ${\vec r^i}$ as observed from the TF, ${\vec \omega ^t}{\rm{ and}}\,\,{\vec \omega ^i}$ are the absolute angular velocity vectors, and ${\vec \alpha ^t}\,\,{\rm{and}}\,\,{\vec \alpha ^i}$ the absolute angular acceleration vectors, of the TF associated with body $i$ and the BF of body $i$, respectively. These vector-equations can be projected to the GF as follows:

\begin{equation} \label{Eq3_16}
\dot{\bf{R}}_P^i = \dot{\bf{R}}^t + {\bf A}^t \dot {\bar {\bf {r}}}^i + {\bf A}^t \left( \tilde{\bar{\omega}}^t \bar{\bf{r}}^i \right) + {\bf A}^i \left( \tilde{\hat{\omega}}^t {\hat{\bf u}}_P^i \right)
\end{equation}

\begin{equation} \label{Eq3_17}
\ddot{{\bf R}}_P^i = \ddot{\bf R}^t + {{\bf{A}}^t}\ddot {\bf\bar r}^i + {{\bf{A}}^t}\left( \tilde{\bar{\alpha}}^t + \tilde{\bar{\omega}}^t\tilde{\bar{\omega}}^t \right){{\bf{\bar r}}^i} + 2{{\bf{A}}^t}\left( \tilde{\bar{\omega}}^t\dot {\bar {\bf {r}}}^i \right) + {{\bf{A}}^i}\left( \tilde{\hat{\alpha}}^i + \tilde{\hat{\omega}}^i \tilde{\hat{\omega}}^i \right) {\bf{\hat u}}_P^i
\end{equation}

where parentheses in Eq. \ref{Eq3_16} are used just for ease of reading. In these equations, symbol “tilde” over a vector, like ${\bf{\tilde b}}$, means the skew-symmetric matrix associated with the column matrix ${\bf{b}}$. 
Because the TF’s velocity and acceleration vectors are more easily projected in the track-frame, Eqs. \ref{Eq3_16} – \ref{Eq3_17} are also projected to that frame pre-multiplying both sides by ${\left( {{{\bf{A}}^t}} \right)^T}$ . The following expressions are easily deduced: 

\begin{equation} \label{Eq3_18}
\bar{ \bf {\dot R}}_P^i = { \bar {\bf {\dot {R}}}^t} + \dot {\bar {\bf {r}}}^i + \tilde{\bar{\omega}}^t {\bar {\bf r}^i} + {{\bf{A}}^{t,i}}\left( \tilde{\hat{\omega}}^i {\hat{\bf u}}_P^i \right)
\end{equation}

\begin{equation} \label{Eq3_19}
\bar{ \bf {\ddot R}}_P^i = \bar{ \bf {\ddot R}}^t + \ddot {\bf {\bar{r}}}^i + \left( \tilde{\bar{\alpha}}^t + \tilde{\bar{\omega}}^t\tilde{\bar{\omega}}^t \right){\bar {\bf r}^i}+ 2\tilde{\bar{\omega}}^t \dot {\bar {\bf {r}}}^i + {{\bf{A}}^{t,i}}\left( \tilde{\hat{\alpha}}^i + \tilde{\hat{\omega}}^i\tilde{\hat{\omega}}^i \right){\bf{\hat u}}_P^i
\end{equation}

In order to compute Eqs. \ref{Eq3_18} - \ref{Eq3_19}, the orientation matrices, angular velocities and angular accelerations of the different frames need to be computed as a function of the generalized coordinates and velocities. The angular velocity ${{\bf{\bar \omega }}^t}$ and acceleration ${{\bf{\bar \alpha }}^t}$ vectors of the TF are given in Eq. \ref{Eq3_5}. The angular velocity of body $i$ with respect to the TF is obtained, under the small-angles assumption, as follows:

\begin{equation} \label{Eq3_20}
{{\bf{\hat \omega }}^{t,i}} = \left[ {\begin{array}{*{20}{c}}
{{{\dot \varphi }^i}}\\
{{{\dot \theta }^i}}\\
{{{\dot \psi }^i}}
\end{array}} \right],\,\,\,\,\,\,{{\bf{\bar \omega }}^{t,i}} = {{\bf{A}}^{t,i}}{{\bf{\hat \omega }}^{t,i}} = \left[ {\begin{array}{*{20}{c}}
1&{{\psi ^i}}&{ - {\theta ^i}}\\
{ - {\psi ^i}}&1&{{\varphi ^i}}\\
{{\theta ^i}}&{ - {\varphi ^i}}&1
\end{array}} \right]\left[ {\begin{array}{*{20}{c}}
{{{\dot \varphi }^i}}\\
{{{\dot \theta }^i}}\\
{{{\dot \psi }^i}}
\end{array}} \right]
\end{equation}

and the absolute angular velocity of body $i$, under the small-angles assumption, is given by: 

\begin{equation} \label{Eq3_21}
{{\bf{\hat \omega }}^i} = {{\bf{\hat \omega }}^t} + {{\bf{\hat \omega }}^{t,i}} = {\left( {{{\bf{A}}^{t,i}}} \right)^T}{{\bf{\bar \omega }}^t} + {{\bf{\hat \omega }}^{t,i}}
\end{equation}

The absolute angular acceleration of body $i$, ${{\bf{\hat \alpha }}^i}$ is simply calculated as the time-derivative of Eq. \ref{Eq3_21}.

\section{Kinematics of the computer vision} \label{sec:kinVis}

\begin{figure}[htbp!]
	\centering
	\includegraphics[width=0.7\linewidth]{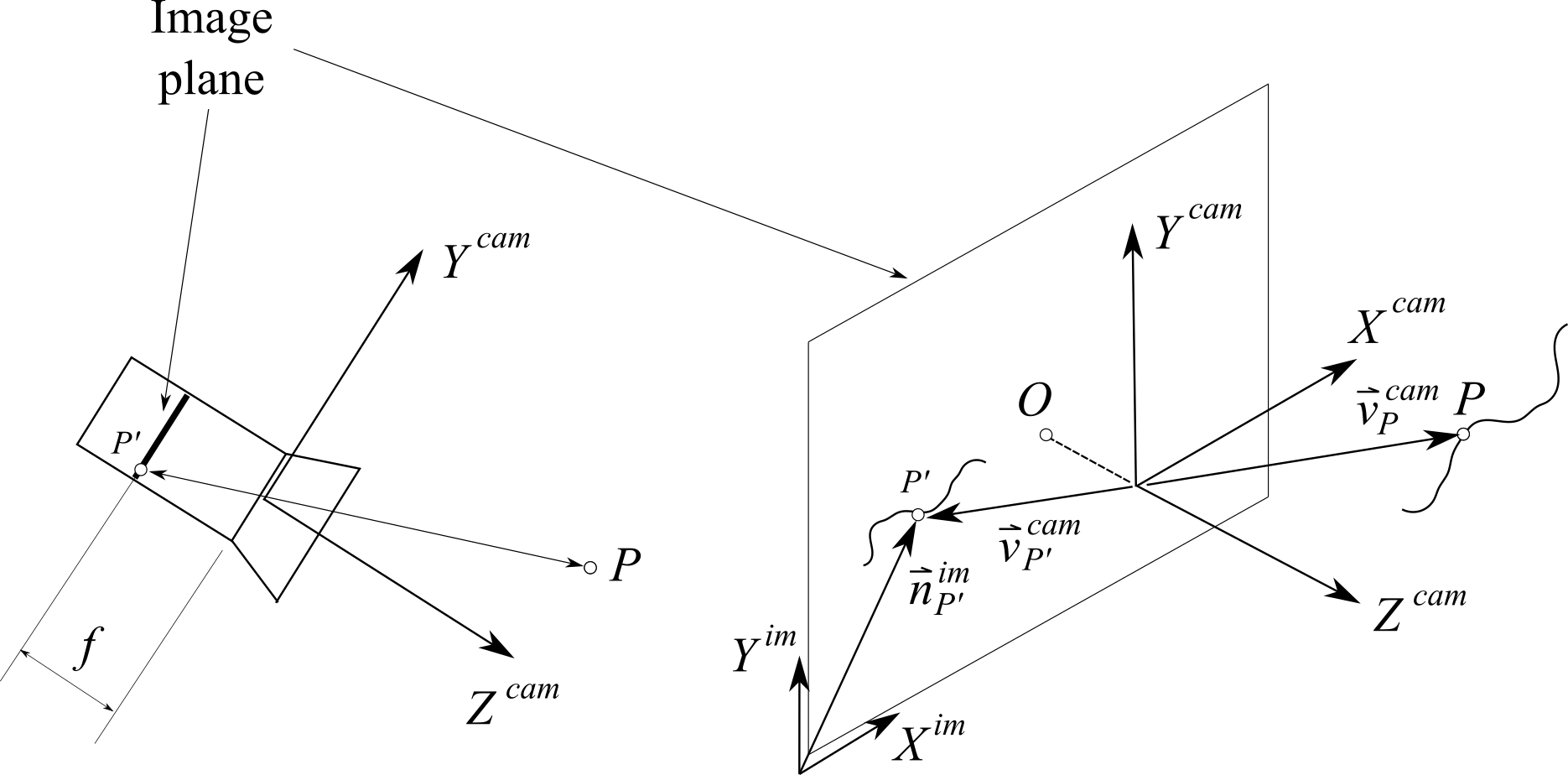}
	\caption{Kinematics of the computer vision}
	\label{fig:fig4_1}
\end{figure}

The kinematics of the computer vision used here is described in detail in \cite{escalona2020computerVision}. Using the pin-hole model of the camera, Fig. \ref{fig:fig4_1} shows the relation between the position vector $\vec n_{P'}^{im}$ of an arbitrary point $P$ in the camera frame $<X^{cam}, Y^{cam}, Z^{cam}>$ (in our problem it can be left cam $lcam$ or right cam $rcam$) and the position vector   of the recorded point P’ in the image plane $<X^{im}, Y^{im}>$.

\begin{figure}[htbp!]
	\centering
	\includegraphics[width=0.4\linewidth]{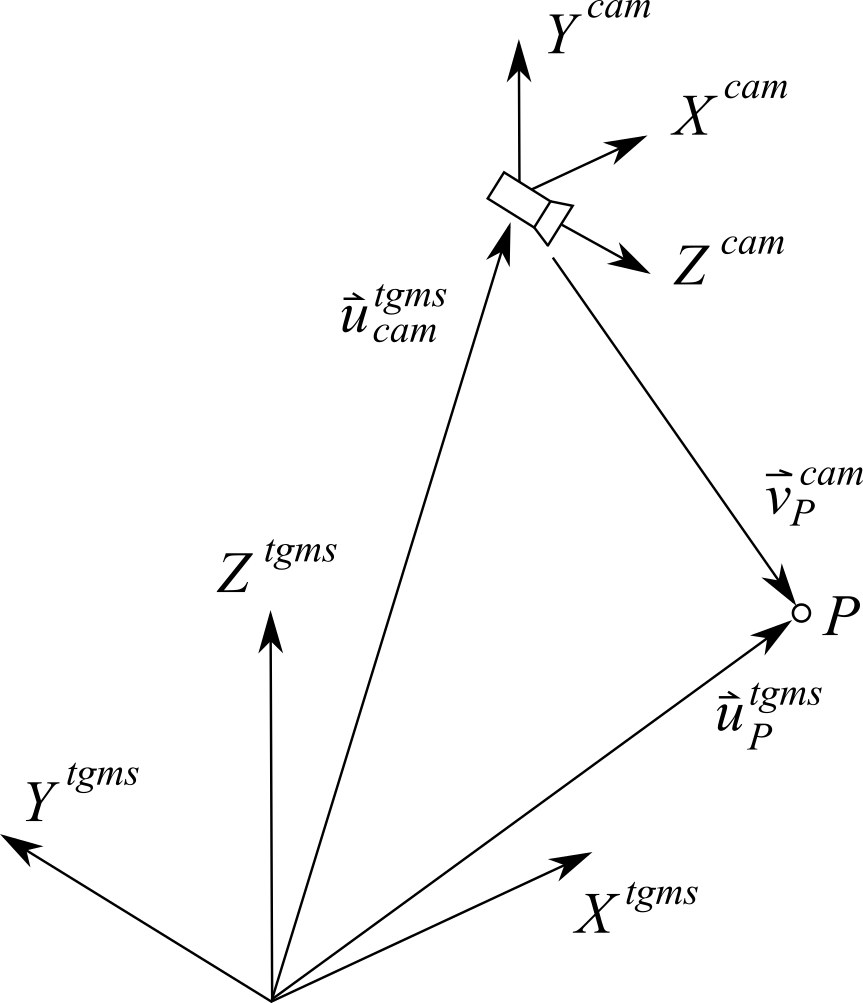}
	\caption{TGMS frame and camera frame}
	\label{fig:fig4_2}
\end{figure}

Figure \ref{fig:fig4_2} shows the location of the camera in the TGMS and the relation between the position vector $\vec v_P^{cam}$  of the arbitrary point $P$ in the camera frame and its position vector $\vec v_P^{tgms}$ in the TGMS frame.

As explained in \cite{escalona2020computerVision}, the components of vectors $\vec n_{P'}^{im}$ and $\vec u_P^{tgms}$ are related through the equation:

\begin{equation} \label{Eq4_1}
c\left[ {\begin{array}{*{20}{c}}
{{\bf{n}}_{P'}^{im}}\\
1
\end{array}} \right] = {{\bf{M}}^{int}}{{\bf{M}}^{ext}}\left[ {\begin{array}{*{20}{c}}
{{\bf{\hat u}}_P^{tgms}}\\
1
\end{array}} \right]
\end{equation}

where $c$ is an unknown $scale$ $factor$. The matrix product on the right-hand side of this equation is called \textit{projection matrix} ${\bf{P}} = {{\bf{M}}^{int}}{{\bf{M}}^{ext}}$ . The column matrix ${\bf{n}}_{P'}^{im}$ is $2 \times 1$ (image is planar), and its components are given in pixel units (dimensionless) while the column matrix ${\bf{\hat u}}_P^{tgms}$ is $3 \times 1$, and its components are given in meters. These dimensions explain that the projection matrix \textbf{P} is $3 \times 4$. Matrix ${{\bf{M}}^{int}}$ is $3 \times 3$ and it is called \textit{matrix of intrinsic parameters} of the camera, and matrix ${{\bf{M}}^{ext}}$ is $3 \times 4$, it is called \textit{matrix of extrinsic parameters} of the camera, and it is given by:

\begin{equation} \label{Eq4_2}
{{\bf{M}}^{ext}} = \left[ {\begin{array}{*{20}{c}}
{{{\left( {{{\bf{A}}^{tgms,cam}}} \right)}^T}}&{ - {{\left( {{{\bf{A}}^{tgms,cam}}} \right)}^T}{\bf{\hat u}}_{cam}^{tgms}}
\end{array}} \right]
\end{equation}

Matrices of intrinsic and extrinsic parameters can be experimentally obtained using the Zhang calibration method \cite{zhang1998camCal} that is also described in \cite{escalona2020computerVision}.

Just using Eq. \ref{Eq4_1}, the position vector ${\bf{\hat u}}_P^{tgms}$ of the point $P$ cannot be obtained using the values of ${\bf{n}}_{P'}^{im}$ because there are 4 unknowns (three components of ${\bf{\hat u}}_P^{tgms}$ and the scale factor $c$). One exception occurs when the point $P$ moves on a surface whose equation is known in the TGMS frame. This is the case at hand if $P$ belongs to the plane highlighted by the laser projector. In this case, the following system of equations can be solved to find ${\bf{\hat u}}_P^{tgms}$: 

\begin{equation} \label{Eq4_3}
\left\{ {\begin{array}{*{20}{c}}
{c\left[ {\begin{array}{*{20}{c}}
{{\bf{n}}_{P'}^{im}}\\
1
\end{array}} \right] = {{\bf{M}}^{int}}{{\bf{M}}^{ext}}\left[ {\begin{array}{*{20}{c}}
{{\bf{\hat u}}_P^{tgms}}\\
1
\end{array}} \right]}\\
{{A^{las}}{{\left[ {u_P^{tgms}} \right]}_x} + {B^{las}}{{\left[ {u_P^{tgms}} \right]}_y} + {C^{las}}{{\left[ {u_P^{tgms}} \right]}_z} + {D^{las}} = 0}
\end{array}} \right.\,
\end{equation}

where ${A^{las}},\,\,{B^{las}},\,\,{C^{las}}$ and ${D^{las}}$  are the constants that define the laser plane (left laser, $llas$, or right laser, $rlas$, in our case), and ${\left[ {u_P^{tgms}} \right]_k}\,\,\,k = x,y,z$ means the component $k$ of the vector $\vec u_P^{tgms}$ in the TGMS frame. The constants that define the laser planes have to be experimentally obtained in the TGMS computer vision-calibration process described in Section \ref{sec:camCal}. Equation \ref{Eq4_3} is a system of 4 algebraic equations with 4 unknowns that can be used to find vector components ${\bf{\hat u}}_P^{tgms}$ using as input data the vector components ${\bf{n}}_{P'}^{im}$  and the parameters of the cameras and the lasers.

\section{Detecting the rail cross-section from a camera frame} \label{sec:detectXsec}

As a result of the solution of Eq. \ref{Eq4_3} for all highlighted pixels in the image fames, a cloud of points $P$ in the right rail and a cloud of points $Q$ in the left rail, with position vectors ${\bf{\hat u}}_P^{tgms}$ and ${\bf{\hat u}}_Q^{tgms}$, respectively, that belong to the rails cross-sections can be identified. In fact, the points do not really belong to cross-sections, just to sections, because the laser planes are not necessarily perpendicular to the rails center line. However, because the relative angles of the TGMS with respect to the TF are very small, the irregularities are also small, and the lasers are set to project the light plane at right angles with respect to the rails, it will be assumed that the highlighted sections are actually cross-sections.

\begin{figure}[htbp!]
	\centering
	\includegraphics[width=0.5\linewidth]{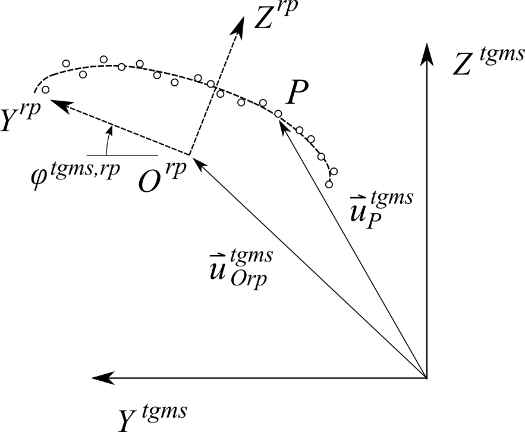}
	\caption{Cloud of points detected with computer vision}
	\label{fig:fig5_1}
\end{figure}

\begin{figure}[htbp!]
	\centering
	\includegraphics[width=0.4\linewidth]{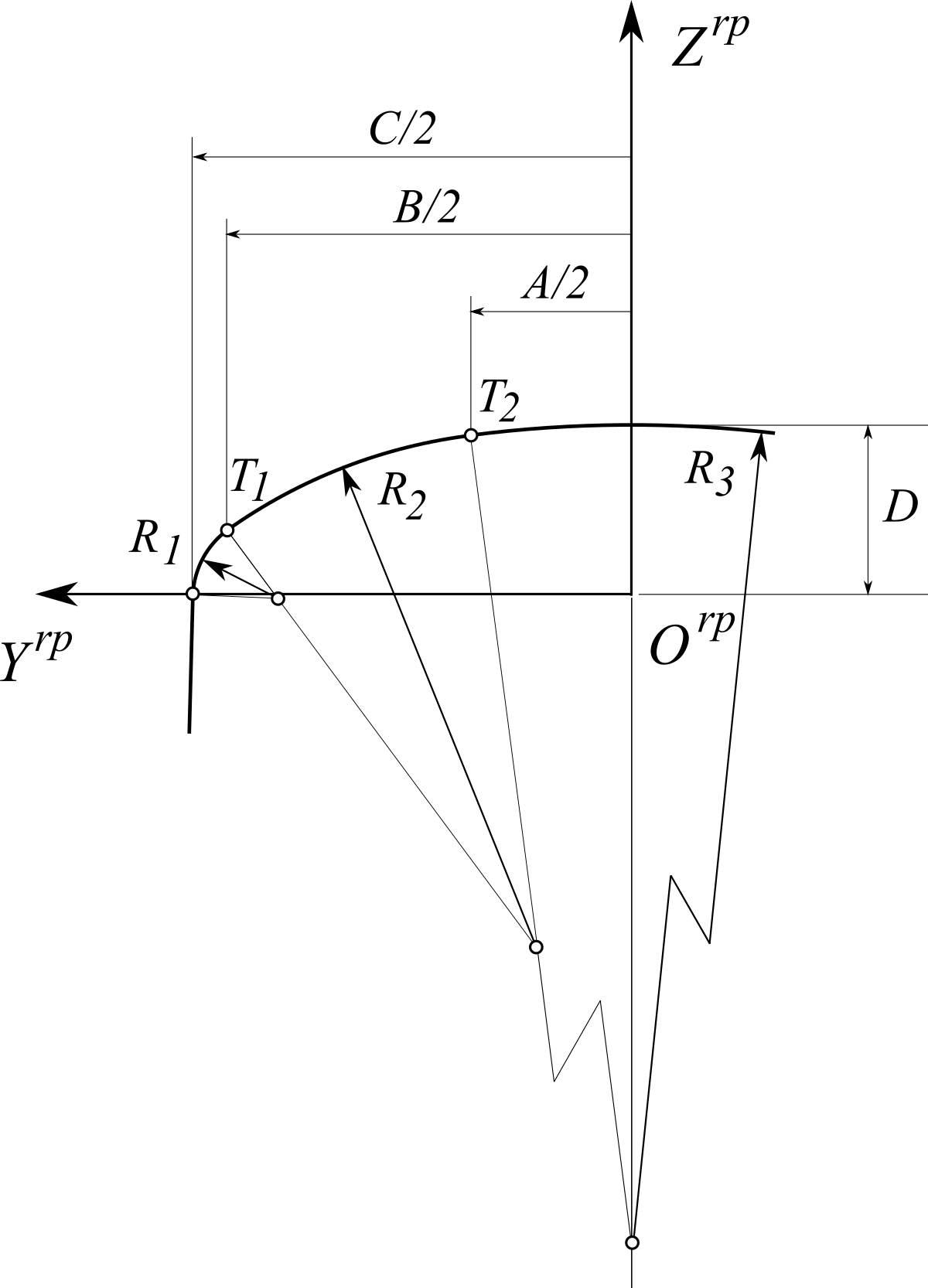}
	\caption{UIC 54 E1 rail-head profile}
	\label{fig:fig5_2}
\end{figure}

Figure \ref{fig:fig5_1} shows a sketch of the cloud of points and, in dashed line, the theoretical rail-head profile. The theoretical rail-head profiles, when they are new, not worn, have a known geometry that is made of circular and straight segments. An example is the UIC 54 E1 rail-head profile shown in Fig. \ref{fig:fig5_2}. Detecting the rail cross-section from a camera frame consists on solving an optimization problem to find the position ${\bf{\hat u}}_{Orp}^{tgms}$  and orientation ${\varphi ^{tgms,rp}}$ of the rail profile that better fits the cloud of points. This will be the assumed position and orientation of the rail head profile ($lrp$ or $rrp$) in the TGMS frame when the vehicle is moving. The optimization process is detailed next.

\begin{figure}[htbp!]
	\centering
	\includegraphics[width=0.5\linewidth]{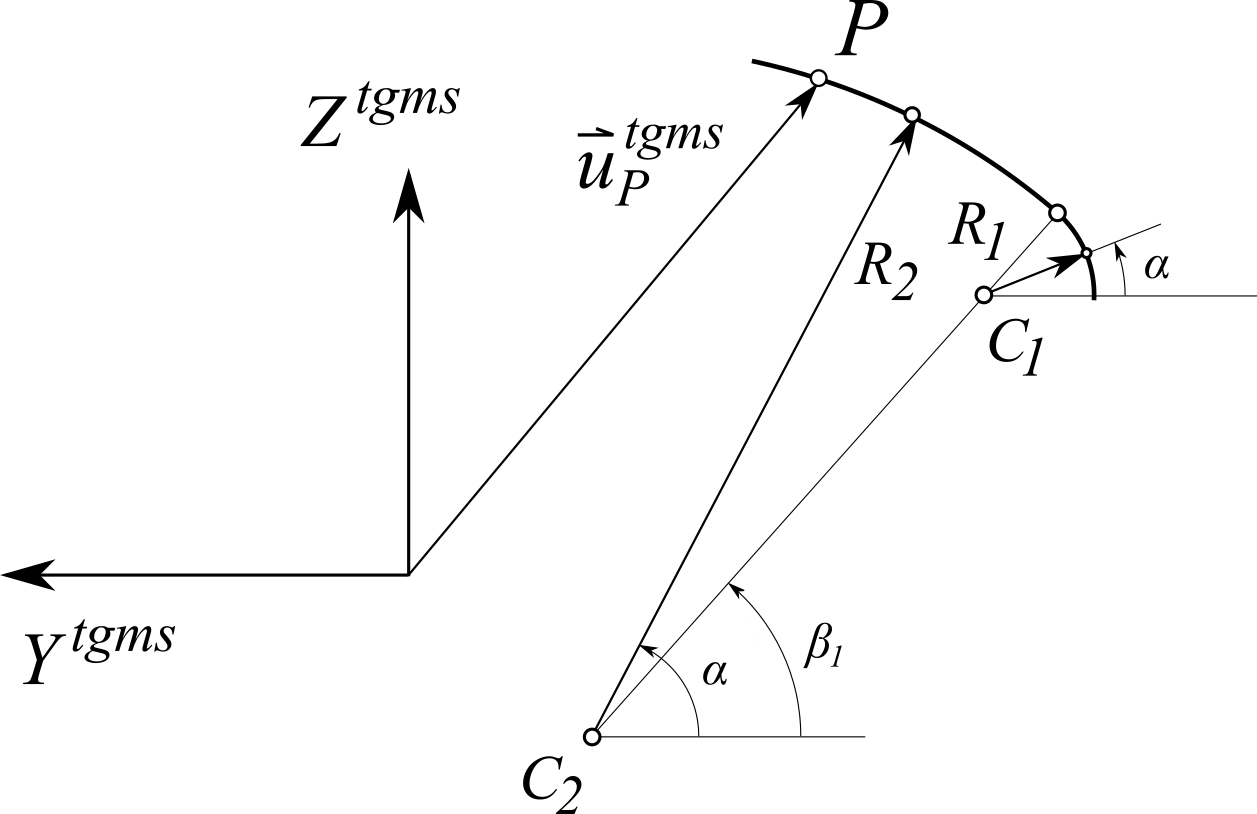}
	\caption{Detection of position and orientation of UIC 54 E1 profile}
	\label{fig:fig5_3}
\end{figure}

The parametric expression of the rail-head profiles is easy to obtain. Figure \ref{fig:fig5_3} shows a profile segment that contains the tangent point $T_1$ between the arcs with radius $R_1$ and $R_2$ (see Fig. \ref{fig:fig5_2}). Using as a parameter of the curve the angle $\alpha$ that is observed in the figure, the parametric equations of the coordinates of a point in the profile are given by:

\begin{equation} \label{Eq5_1}
{{\bf{\hat u}}^{tgms}}\left( \alpha  \right) = \left\{ {\begin{array}{*{20}{c}}
{{{{\bf{\hat u}}}_{C1}} + {R_1}{{\left[ {\begin{array}{*{20}{c}}
{\cos \alpha }&{\sin \alpha }
\end{array}} \right]}^T}\,\,{\rm{  if }}\,\alpha  < {\beta _{\rm{1}}}}\\
{{{{\bf{\hat u}}}_{C2}} + {R_2}{{\left[ {\begin{array}{*{20}{c}}
{\cos \alpha }&{\sin \alpha }
\end{array}} \right]}^T}\,\,{\rm{  if }}\,\alpha  > {\beta _{\rm{1}}}}
\end{array}} \right.
\end{equation}

where the position vectors of the centers of the circles  $C_1$ and $C_2$ and the angle $\beta_1$are given by:

\begin{equation} \label{Eq5_2}
{{\bf{\hat u}}_{C1}} = {\left[ {\begin{array}{*{20}{c}}
{{y_{C1}}}&{{z_{C1}}}
\end{array}} \right]^T},\,\,\,{{\bf{\hat u}}_{C2}} = {\left[ {\begin{array}{*{20}{c}}
{{y_{C2}}}&{{z_{C2}}}
\end{array}} \right]^T},\,\,\,{\beta _{\rm{1}}} = ta{n^{ - 1}}\left( {\frac{{{z_{C1}} - {z_{C2}}}}{{{y_{C1}} - {y_{C2}}}}} \right)
\end{equation}

The optimization procedure that provides the position and orientation of the rail profile is based on the minimization of the sum of the squared distances from the cloud of observed points to the theoretical profile, this is, a \textit{least squares fit}. The squared distance from a particular point $i$ of the cloud to the theoretical profile given in Eq. \ref{Eq5_1} is given by:

\begin{equation} \label{Eq5_3}
d_i^2 = {\left[ {{\bf{\hat u}}_i^{tgms} - {{{\bf{\hat u}}}^{tgms}}\left( {{\alpha _i}} \right)} \right]^T}\left[ {{\bf{\hat u}}_i^{tgms} - {{{\bf{\hat u}}}^{tgms}}\left( {{\alpha _i}} \right)} \right]
\end{equation}

where ${\bf{\hat u}}_i^{tgms}$  is the position vector of point $i$ of the cloud and ${{\bf{\hat u}}^{tgms}}\left( {{\alpha _i}} \right)$ is the corresponding position in the theoretical profile that is evaluated with Eq. \ref{Eq5_1}. In this equation, the angular parameter $\alpha_i$ associated with point $i$ is obtained as:

\begin{equation} \label{Eq5_4}
{\alpha _i} = \left\{ {\begin{array}{*{20}{c}}
{ta{n^{ - 1}}\left( {\frac{{{z_i} - {z_{C1}}}}{{{y_i} - {y_{C1}}}}} \right)\,{\rm{  if }}\,{\alpha _i} < {\beta _{\rm{1}}}}\\
{ta{n^{ - 1}}\left( {\frac{{{z_i} - {z_{C2}}}}{{{y_i} - {y_{C2}}}}} \right)\,{\rm{  if }}\,{\alpha _i} > {\beta _{\rm{1}}}}
\end{array}} \right.
\end{equation}

Therefore, the least squares fitting is the result of the minimization of the following function: 

\begin{equation} \label{Eq5_5}
\begin{array}{l}
f\left( {\bf{x}} \right) = \sum\limits_{i = 1}^{np} {{{\left[ {{\bf{\hat u}}_i^{tgms} - {{{\bf{\hat u}}}^{tgms}}\left( {{\alpha _i}} \right)} \right]}^T}\left[ {{\bf{\hat u}}_i^{tgms} - {{{\bf{\hat u}}}^{tgms}}\left( {{\alpha _i}} \right)} \right]} \\
{\bf{x}} = {\left[ {\begin{array}{*{20}{c}}
{{y_{C1}}}&{{z_{C1}}}&{{y_{C2}}}&{{z_{C2}}}
\end{array}} \right]^T}
\end{array}
\end{equation}

where $np$ is the number of points in the cloud. Equation \ref{Eq5_5} can be used to find, using as an input the position vector of the points in the cloud, the position vector of the centers of the circles $C_1$ and $C_2$ in plane $<Y^{tgms}, Z^{tgms}>$. The radii $R_1$ and $R_2$ are input data.

The minimization of the function shown in Eq. \ref{Eq5_5} is a problem of constrained minimization.  An algebraic constraint is needed to guarantee that the distance between the centers $C_1$ and $C_2$ equals the radius difference $R_2 - R_1$, otherwise, the two circular arcs would not be connected at $T_1$. The algebraic constraint yields:

\begin{equation} \label{Eq5_6}
g\left( {\bf{x}} \right) = {\left( {{y_{C1}} - {y_{C2}}} \right)^2} + {\left( {{z_{C1}} - {z_{C2}}} \right)^2} - {\left( {{R_2} - {R_1}} \right)^2} = 0
\end{equation}

Therefore, the profile fitting is the result of solving the problem: 

\begin{equation} \label{Eq5_7}
\begin{array}{l}
\mathop {\min }\limits_{\bf{x}} f\left( {\bf{x}} \right)\\
{\rm{subjected} \: \rm{to} \:}g\left( {\bf{x}} \right) = 0
\end{array}
\end{equation}

Using the method of Lagrange multipliers, the minimization problem is equivalent to solve the following 5 non-linear algebraic equations:

\begin{equation} \label{Eq5_8}
\begin{array}{l}
\frac{{\partial f\left( {\bf{x}} \right)}}{{\partial {\bf{x}}}} + \lambda \frac{{\partial g\left( {\bf{x}} \right)}}{{\partial {\bf{x}}}} = 0\\
g\left( {\bf{x}} \right) = 0
\end{array}
\end{equation}

where the unknowns are the 4 components of ${\bf{x}} = {\left[ {\begin{array}{*{20}{c}}
{{y_{C1}}}&{{z_{C1}}}&{{y_{C2}}}&{{z_{C2}}}
\end{array}} \right]^T}$  and the Lagrange multiplier $\lambda$. The position vector ${\bf{\hat u}}_{Orp}^{tgms}$  and orientation ${\varphi ^{tgms,rp}}$ of the rail profile are  easily obtained using simple rigid body kinematics once the position of the centers $C_1$ and $C_2$ is obtained.
This optimization procedure can be easily extended to the whole rail-head profile (5 circles) or any other sub-profile including 3 or 4 circles. The correct method would depend on the segment of the rail-head profile that the laser actually highlights. Moreover, this procedure can be extended to a 3D curve fitting just including the x-component of the position vectors in the cloud of points. Extending the optimization to 3D could result in a better accuracy, but, for sure, it would increase significantly the computational time.

\section{Equations for geometry measurement} \label{sec:geomMeas}

\begin{figure}[htbp!]
	\centering
	\includegraphics[width=0.6\linewidth]{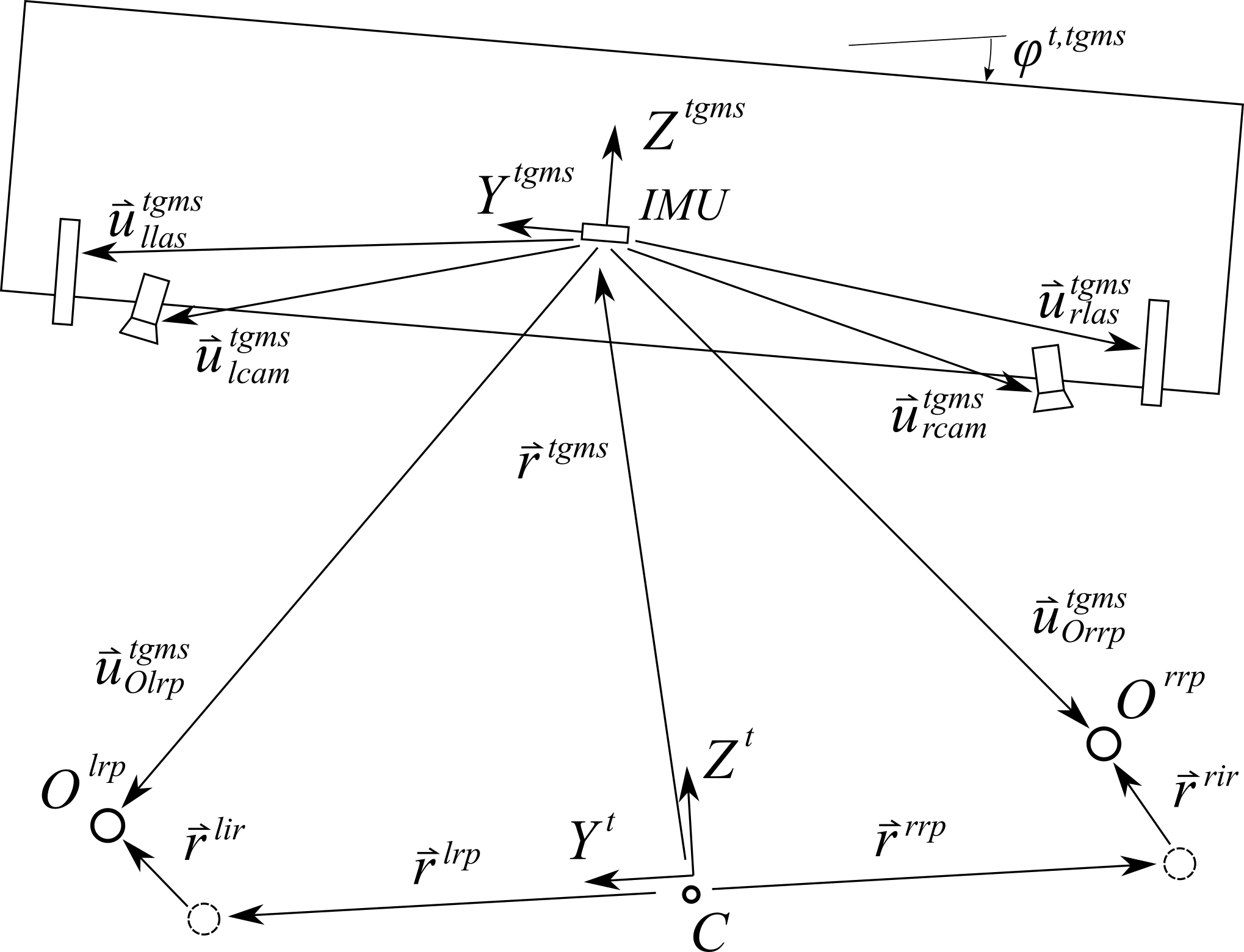}
	\caption{Planar view of the TGMS}
	\label{fig:fig6_1}
\end{figure}

The equations that can be used to measure the track irregularities are easily deduced with the help of Fig. \ref{fig:fig2_2} that for convenience is presented again in Fig. \ref{fig:fig6_1}. In this figure, vectors ${\bf{\hat u}}_{Olrp}^{tgms}\,\,{\rm{and}}\,\,{\bf{\hat u}}_{Orrp}^{tgms}$ are input data from the computer vision algorithm explained in previous section. The following equalities can be easily identified with the help of the figure:

\begin{equation} \label{Eq6_1}
\begin{array}{l}
{{\vec r}^{tgms}} + \vec u_{Olrp}^{tgms} = {{\vec r}^{lrp}} + {{\vec r}^{lir}}\\
{{\vec r}^{tgms}} + \vec u_{Orrp}^{tgms} = {{\vec r}^{rrp}} + {{\vec r}^{rir}}
\end{array}
\end{equation}

Subtracting both vector equations one gets:

\begin{equation} \label{Eq6_2}
\vec u_{Olrp}^{tgms} - \vec u_{Orrp}^{tgms} = {\vec r^{lrp}} + {\vec r^{lir}} - \left( {{{\vec r}^{rrp}} + {{\vec r}^{rir}}} \right)
\end{equation}

In this equation, the position vector ${\vec r^{tgms}}$ of the TGMS does not appear. This vector equation can be projected in the TF, as follows:

\begin{equation} \label{Eq6_3}
{{\bf{A}}^{t,tgms}}\left( {{\bf{\hat u}}_{Olrp}^{tgms} - {\bf{\hat u}}_{Orrp}^{tgms}} \right) = {{\bf{\bar r}}^{lrp}} - {{\bf{\bar r}}^{rrp}} + {{\bf{\bar r}}^{lir}} - {{\bf{\bar r}}^{rir}}
\end{equation}

Using again the small-angles assumption, the $Y, Z$ components of this equation are given by:

\begin{equation} \label{Eq6_4}
\left[ {\begin{array}{*{20}{c}}
1&{ - {\varphi ^{t,tgms}}}\\
{{\varphi ^{t,tgms}}}&1
\end{array}} \right]\left[ {\begin{array}{*{20}{c}}
{{{\left[ {{\bf{\hat u}}_{Olrp}^{tgms}} \right]}_y} - {{\left[ {{\bf{\hat u}}_{Orrp}^{tgms}} \right]}_y}}\\
{{{\left[ {{\bf{\hat u}}_{Olrp}^{tgms}} \right]}_z} - {{\left[ {{\bf{\hat u}}_{Orrp}^{tgms}} \right]}_z}}
\end{array}} \right] = \left[ {\begin{array}{*{20}{c}}
{2{L^r}}\\
0
\end{array}} \right] + \left[ {\begin{array}{*{20}{c}}
{r_y^{lir} - r_y^{rir}}\\
{r_z^{lir} - r_z^{rir}}
\end{array}} \right]
\end{equation}

where $L^r$ is half the distance between the rail-head profiles without irregularities. In this equation,  the result ${{\bf{\bar r}}^{lrp}} - {{\bf{\bar r}}^{rrp}} = {\left[ {\begin{array}{*{20}{c}}
{2{L^r}}&0
\end{array}} \right]^T}$ has been used. According to the definition given in Section \ref{sec:irregularTrack}, the components of the last column matrix of Eq. \ref{Eq6_4} are the gauge variation ($gv$) and the cross-level ($cl$). Therefore, rearranging Eq. \ref{Eq6_4} yields: 

\begin{equation} \label{Eq6_5}
\begin{array}{l}
gv = \left( {{{\left[ {{\bf{\hat u}}_{Olrp}^{tgms}} \right]}_y} - {{\left[ {{\bf{\hat u}}_{Orrp}^{tgms}} \right]}_y}} \right) - {\varphi ^{t,tgms}}\left( {{{\left[ {{\bf{\hat u}}_{Olrp}^{tgms}} \right]}_z} - {{\left[ {{\bf{\hat u}}_{Orrp}^{tgms}} \right]}_z}} \right) - 2{L^r}\\
cl = {\varphi ^{t,tgms}}\left( {{{\left[ {{\bf{\hat u}}_{Olrp}^{tgms}} \right]}_y} - {{\left[ {{\bf{\hat u}}_{Orrp}^{tgms}} \right]}_y}} \right) + \left( {{{\left[ {{\bf{\hat u}}_{Olrp}^{tgms}} \right]}_z} - {{\left[ {{\bf{\hat u}}_{Orrp}^{tgms}} \right]}_z}} \right)
\end{array}
\end{equation}

Adding the vector equations Eq. \ref{Eq6_1}, one gets: 

\begin{equation} \label{Eq6_6}
2{\vec r^{tgms}} + \vec u_Q^{tgms} + \vec u_P^{tgms} = {\vec r^{lir}} + {\vec r^{rir}}
\end{equation}

where the fact that ${\vec r^{lrp}} + {\vec r^{rrp}} = \vec 0$ has been used. Using again the small-angles assumption, the $Y, Z$ components of this equation are given by:

\begin{equation} \label{Eq6_7}
2\left[ {\begin{array}{*{20}{c}}
{r_y^{tgms}}\\
{r_z^{tgms}}
\end{array}} \right] + \left[ {\begin{array}{*{20}{c}}
1&{ - {\varphi ^{t,tgms}}}\\
{{\varphi ^{t,tgms}}}&1
\end{array}} \right]\left[ {\begin{array}{*{20}{c}}
{{{\left[ {{\bf{\hat u}}_{Olrp}^{tgms}} \right]}_y} + {{\left[ {{\bf{\hat u}}_{Orrp}^{tgms}} \right]}_y}}\\
{{{\left[ {{\bf{\hat u}}_{Olrp}^{tgms}} \right]}_z} + {{\left[ {{\bf{\hat u}}_{Orrp}^{tgms}} \right]}_z}}
\end{array}} \right] = \left[ {\begin{array}{*{20}{c}}
{r_y^{lir} + r_y^{rir}}\\
{r_z^{lir} + r_z^{rir}}
\end{array}} \right]
\end{equation}

According to the definition given in Section \ref{sec:irregularTrack}, the components of the last column matrix of Eq. \ref{Eq6_7} are twice the alignment irregularity ($al$) and twice the vertical profile ($vp$). Therefore, rearranging Eq. \ref{Eq5_7} yields:

\begin{equation} \label{Eq6_8}
\begin{array}{l}
al = \frac{1}{2}\left( {{{\left[ {{\bf{\hat u}}_{Olrp}^{tgms}} \right]}_y} + {{\left[ {{\bf{\hat u}}_{Orrp}^{tgms}} \right]}_y}} \right) - \frac{{{\varphi ^{t,tgms}}}}{2}\left( {{{\left[ {{\bf{\hat u}}_{Olrp}^{tgms}} \right]}_z} + {{\left[ {{\bf{\hat u}}_{Orrp}^{tgms}} \right]}_z}} \right) + r_y^{tgms}\\
vp = \frac{{{\varphi ^{t,tgms}}}}{2}\left( {{{\left[ {{\bf{\hat u}}_{Olrp}^{tgms}} \right]}_y} + {{\left[ {{\bf{\hat u}}_{Orrp}^{tgms}} \right]}_y}} \right) + \frac{1}{2}\left( {{{\left[ {{\bf{\hat u}}_{Olrp}^{tgms}} \right]}_z} + {{\left[ {{\bf{\hat u}}_{Orrp}^{tgms}} \right]}_z}} \right) + r_z^{tgms}
\end{array}
\end{equation}

Therefore, Eqs. \ref{Eq6_5} and \ref{Eq6_8} can be used to find all track irregularities. The following conclusions are highlighted:

\begin{enumerate}
    \item The calculation of the relative track irregularities ($gv$ and $cl$), as shown in Eq. \ref{Eq6_5}, needs as an input the output of the computer vision ${\bf{\hat u}}_{Olrp}^{tgms}\,\,{\rm{and}}\,\,{\bf{\hat u}}_{Orrp}^{tgms}$ and the roll angle of the TGMS with respect to the track ${\varphi ^{t,tgms}}$. 
    \item The calculation of the absolute track irregularities ($al$ and $vp$), as shown in Eq. \ref{Eq6_8}, needs, in addition, the relative trajectory ${{\bf{\bar r}}^{tgms}}$ of the TGMS with respect to the TF.
\end{enumerate}

\section{Measurement of TGMS to TF relative motion} \label{sec:relMot}

As explained in previous section, the relative trajectory ${{\bf{\bar r}}^{tgms}}$  of the TGMS with respect to the TF is needed to find the absolute irregularities of the track using Eq. \ref{Eq6_8}. This is not an easy task. Because ${{\bf{\bar r}}^{tgms}}$ is a relative motion, finding it requires information about the instantaneous position of the TGMS frame and the instantaneous position and orientation of the TF. This is obtained as follows: 

\begin{enumerate}
    \item The IMU provides information about the absolute angular velocity and the absolute acceleration of the TGMS. The accelerometer data are three (noisy) signals that measure the following vector components:
    
\begin{equation} \label{Eq7_1}
{{\bf{a}}^{imu}} = \hat{ \bf {\ddot R}}^{tgms} + {\left( {{{\bf{A}}^{tgms}}} \right)^T}{\left[ {\begin{array}{*{20}{c}}
0&0&g
\end{array}} \right]^T}
\end{equation}

this is, the absolute acceleration in the sensor frame, plus the gravitational constant $g$, that is assumed to act in the absolute $Z$ direction. The gravity field is added to the accelerometer signals because, in general, IMUs use capacitive accelerometers.
    \item The information about the TF position and orientation is obtained from the ideal geometry of the track that is provided by the track pre-processor. To that end, the position of the TGMS along the track ${s^{tgms}}$ at any instant has to be obtained as the entry to the track pre-processor. If an accurate input value of ${s^{tgms}}$ is not available, the value of ${s^{tgms}}$ can be obtained in two phases: a first phase to find an approximate value and a second phase to find a refined value, as follows:
    
    \begin{enumerate}
        \item The $approximate$ $s_{app}^{tgms}$ can be obtained with the help of an encoder that registers the rolling rotation of one wheel of the train. Assuming $rolling-without-slipping$, the position of the TGMS along the track is obtained. However, the result is inaccurate and it drifts with time, because rolling-without-slipping is just an approximation (wheel sliding occurs and micro-slip in the contact patch is the usual situation) and because the rolling radius of the wheel is not known and it varies with the lateral position of the wheel and wear.
        \item The $refined$ $s_{ref}^{tgms}$  is obtained as the output of an algorithm called $odometry$ $algorithm$ that is explained in Section \ref{sec:odometry}. The odometry algorithm refines the ${s^{tgms}}$ signal detecting the instants when the TGMS enters the curves of the track. Because the location of the curves is known, these values are used to correct the approximate ${s^{tgms}}$ signal.
    \end{enumerate}
        
\end{enumerate}

All these calculations are explained in detail in this section and the followings.

\begin{figure}[htbp!]
	\centering
	\includegraphics[width=0.6\linewidth]{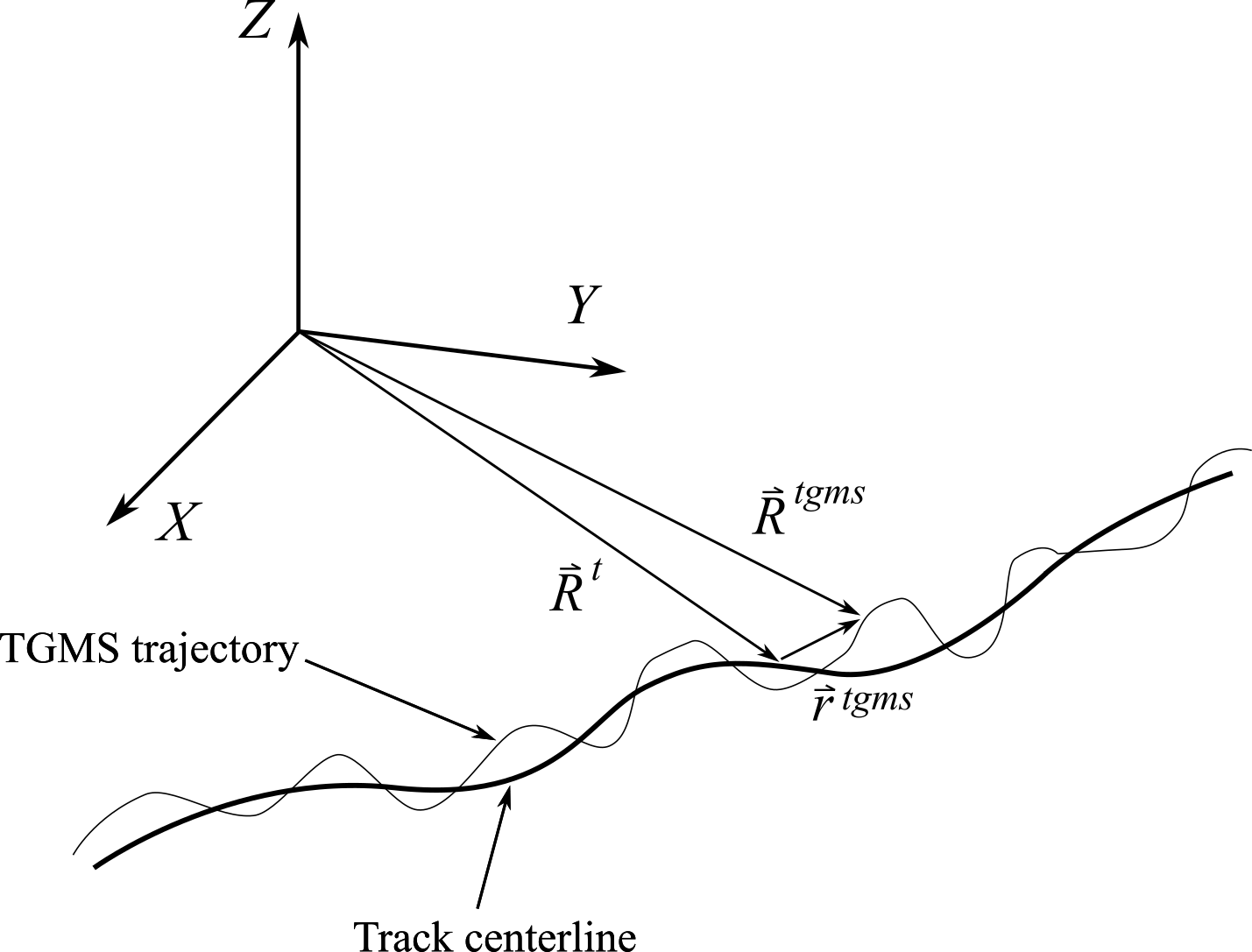}
	\caption{TGMS trajectory}
	\label{fig:fig7_1}
\end{figure}

As it can be observed in Fig. \ref{fig:fig7_1}, the trajectory followed by the TGMS when the vehicle is moving is a 3D curve that slightly varies with respect to the track centerline. In fact, the difference between these 3D curves is what it is needed to measure the absolute track irregularities. 

The absolute acceleration of the TGMS can be obtained using Eq. \ref{Eq3_19} and setting ${\bf{\hat u}}_P^i = {\bf{0}}$ , as follows:

\begin{equation} \label{Eq7_2}
{\bar {\bf {\ddot  {R}}}^{tgms}} = {\hat {\bf {\ddot  {R}}}^{t}} + {\ddot {\bf {\bar {r}}}}^{tgms} + \left( {{\bf{\tilde {\bar {\alpha}} }}}^t + {\bf{\tilde {\bar {\omega }}}^t}{\bf{\tilde {\bar {\omega }}}^t} \right){\bf{\bar r}}^{tgms} + 2{\tilde {\bar{\omega }}^t} {\dot {\bf {\bar r}}}^{tgms}
\end{equation}

This equation has the following scalar components: 

$
{\bar {\bf {\ddot  {R}}}^{tgms}} = \left[ {\begin{array}{*{20}{c}}
{\dot V}\\
{{\rho _h}{V^2}}\\
{ - {\rho _v}{V^2}}
\end{array}} \right] + \left[ {\begin{array}{*{20}{c}}
0\\
{\ddot r_y^{tgms}}\\
{\ddot r_z^{tgms}}
\end{array}} \right] + \left[ {\begin{array}{*{20}{c}}
{ - r_y^{tgms}\left( {\dot V{\rho _h} + {V^2}\left( {{{\rho '}_h} - {\rho _{tw}}{\rho _v}} \right)} \right) + r_z^{tgms}\left( {{V^2}{\rho _{tw}}{\rho _h} + \dot V{\rho _v}} \right)}\\
{ - r_y^{tgms}\left( {{V^2}\left( {\rho _{tw}^2 + \rho _h^2} \right)} \right) - r_z^{tgms}\left( { - {V^2}{\rho _v}{\rho _h} + \dot V{\rho _{tw}}} \right)}\\
{r_y^{tgms}\left( {{V^2}{\rho _v}{\rho _h} + \dot V{\rho _{tw}}} \right) - r_z^{tgms}\left( {{V^2}\left( {\rho _{tw}^2 + \rho _v^2} \right)} \right)}
\end{array}} \right] +
$

\begin{equation} \label{Eq7_3}
+ 2\left[ {\begin{array}{*{20}{c}}
V\dot{r}_z^{tgms}{\rho _v} - V\dot{r}_y^{tgms}\rho _h\\
 - V\dot{r}_z^{tgms}\rho _{tw}\\
V\dot {r}_y^{tgms}\rho _{tw}
\end{array}} \right]
\end{equation}

where the expressions:

\begin{equation} \label{Eq7_4}
{{\bf{\bar r}}^{tgms}} = \left[ {\begin{array}{*{20}{c}}
0\\
{r_y^{tgms}}\\
{r_y^{tgms}}
\end{array}} \right],\,\,\,\,{{\bf{\bar \omega }}^t} = \left[ {\begin{array}{*{20}{c}}
{{\rho _{tw}}V}\\
{{\rho _v}V}\\
{{\rho _h}V}
\end{array}} \right],\,\,\,\,\,{{\bf{\bar \alpha }}^t} = \left[ {\begin{array}{*{20}{c}}
{{\rho _{tw}}\dot V}\\
{{\rho _v}\dot V}\\
{{\rho _h}\dot V + {{\rho '}_h}{V^2}}
\end{array}} \right]
\end{equation}

have been substituted in Eq. \ref{Eq7_2} to get Eq. \ref{Eq7_3}. The measure of the accelerometer given in Eq. \ref{Eq7_1} can be projected to the TF, as follows:

${{\bf{A}}^{t,tgms}}{{\bf{a}}^{imu}} = {{\bf{A}}^{t,tgms}}{\hat {\bf {\ddot  {R}}}^{tgms}}  + {{\bf{A}}^{t,tgms}}{\left( {{{\bf{A}}^{tgms}}} \right)^T}{\left[ {\begin{array}{*{20}{c}}
0&0&g
\end{array}} \right]^T} = {\bar {\bf {\ddot  {R}}}^{tgms}}  + {\left( {{{\bf{A}}^t}} \right)^T}{\left[ {\begin{array}{*{20}{c}}
0&0&g
\end{array}} \right]^T} \Rightarrow $  

\begin{equation} \label{Eq7_5}
 \Rightarrow {\bar {\bf {\ddot  {R}}}^{tgms}} = {{\bf{A}}^{t,tgms}}{{\bf{a}}^{imu}} - {\left( {{{\bf{A}}^t}} \right)^T}{\left[ {\begin{array}{*{20}{c}}
0&0&g
\end{array}} \right]^T} = \left[ {\begin{array}{*{20}{c}}
{a_x^{imu} - a_y^{imu}{\psi ^{tgms}} + a_z^{imu}{\theta ^{tgms}}}\\
{a_y^{imu} + a_x^{imu}{\psi ^{tgms}} - a_z^{imu}{\varphi ^{tgms}}}\\
{a_z^{imu} - a_x^{imu}{\theta ^{tgms}} + a_y^{imu}{\varphi ^{tgms}}}
\end{array}} \right] + \left[ {\begin{array}{*{20}{c}}
{g{\theta ^t}}\\
{ - g{\varphi ^t}}\\
{ - g}
\end{array}} \right]
\end{equation}

The second line of Eq. \ref{Eq7_5} equals Eq. \ref{Eq7_3}. Equating the second and third components of these equations and rearranging yields:

\begin{eqnarray} \label{Eq7_6}
\left[ {\begin{array}{*{20}{c}}
{\ddot r_y^{tgms}}\\
{\ddot r_y^{tgms}}
\end{array}} \right] + \left[ {\begin{array}{*{20}{c}}
0&{ - 2V{\rho _{tw}}}\\
{2V{\rho _{tw}}}&0
\end{array}} \right]\left[ {\begin{array}{*{20}{c}}
{\dot r_y^{tgms}}\\
{\dot r_z^{tgms}}
\end{array}} \right] + \left[ {\begin{array}{*{20}{c}}
{ - {V^2}\left( {\rho _{tw}^2 + \rho _h^2} \right)}&{{V^2}{\rho _v}{\rho _h} - \dot V{\rho _{tw}}}\\
{{V^2}{\rho _v}{\rho _h} + \dot V{\rho _{tw}}}&{ - {V^2}\left( {\rho _{tw}^2 + \rho _v^2} \right)}
\end{array}} \right]\left[ {\begin{array}{*{20}{c}}
{r_y^{tgms}}\\
{r_z^{tgms}}
\end{array}} \right] =   \nonumber\\
 = \left[ {\begin{array}{*{20}{c}}
{a_y^{imu} + a_x^{imu}{\psi ^{tgms}} - a_z^{imu}{\varphi ^{tgms}} - g{\varphi ^t} - {\rho _h}{V^2}}\\
{a_z^{imu} - a_x^{imu}{\theta ^{tgms}} + a_y^{imu}{\varphi ^{tgms}} - g + {\rho _v}{V^2}}
\end{array}} \right]
\end{eqnarray}

This is a $2^{nd}$ order linear system of ordinary differential equations (ODE) with time-variant coefficients (\textit{linear time-varying} system, LTV). This ODE has to be integrated forward in time to find the TGMS to TF relative trajectory ($r_y^{tgms}\left( t \right)\,\,{\rm{and}}\,\,r_z^{tgms}\left( t \right)$). The inputs of these equations are:

\begin{enumerate}
    \item The accelerometer data $\bf{a}^{\it{imu}}$.
    \item The instantaneous forward velocity $V$  and acceleration $\dot V$ of the vehicle. This is obtained from the encoder data.
    \item	The position $s^{tgms}$ of the TGMS along the track. This is the output of the odometry algorithm explained in next section. The position $s^{tgms}$ is used as an entry to the track preprocessor to get the track design cant angle ${\varphi ^t}$ and the curvatures ${\rho _{tw}},\,\,{\rho _v}\,{\rm{and}}\,\,{\rho _{tw}}$.
\item	The relative orientation of the TGMS with respect to the TF. This is provided by the set of Euler angles ${\left[ {\begin{array}{*{20}{c}}
{{\varphi ^{tgms}}}&{{\theta ^{tgms}}}&{{\psi ^{tgms}}}
\end{array}} \right]^T}$ . These angles can be obtained using a \textit{sensor fusion algorithm}, as the Madgwick \cite{madgwick2011sensFus} algorithm, that is based on the combination of the gyroscope and the accelerometer data.

\end{enumerate}

In the case of a tangent (straight) track, where all track curvatures are zero, Eq. \ref{Eq7_6} reduces to:

\begin{equation} \label{Eq7_7}
\left[ {\begin{array}{*{20}{c}}
{\ddot r_y^{tgms}}\\
{\ddot r_y^{tgms}}
\end{array}} \right] = \left[ {\begin{array}{*{20}{c}}
{a_y^{imu} + a_x^{imu}{\psi ^{tgms}} - a_z^{imu}{\varphi ^{tgms}}}\\
{a_z^{imu} - a_x^{imu}{\theta ^{tgms}} + a_y^{imu}{\varphi ^{tgms}} - g}
\end{array}} \right]
\end{equation}

Calculation of $V$  and $\dot V$ is a simple task of numerical differentiation of the $s^{tgms}$ signal. The following sub-sections explain the odometry algorithm used to find $s^{tgms}$ and the sensor fusion algorithm used to find ${\left[ {\begin{array}{*{20}{c}}
{{\varphi ^{tgms}}}&{{\theta ^{tgms}}}&{{\psi ^{tgms}}}
\end{array}} \right]^T}$.

\section{Odometry algorithm} \label{sec:odometry}

The odometry algorithm presented here can be used when the TGMS has no access to the data of a precise odometer of the vehicle and/or a GNSS cannot be used, for example, as it happens in underground trains. Underground trains use to be metropolitan. Being metropolitan, there use to be many curved sections. Curved sections facilitate the method presented next.

\begin{figure}[htbp!]
	\centering
	\includegraphics[width=0.7\linewidth]{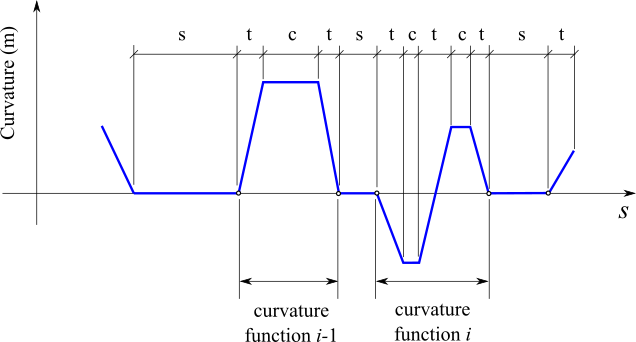}
	\caption{Ideal horizontal curvature of a railway track}
	\label{fig:fig8_1}
\end{figure}

As shown in Fig. \ref{fig:fig8_1}, the ideal geometry of a railway track (horizontal profile, as explained in Section \ref{sec:centerline}) is a succession of segments of three types: straight ($s$ in the figure) with zero curvature, circular ($c$ in the figure) with constant curvature and transitions ($t$ in the figure) with linearly varying curvature. The curvature function can be decomposed into a set of zero segments (straight segments) plus a set of \textit{curvature functions} that have trapezoidal shape (normal curve) or double-trapezoidal shape (\textit{S-curve}). The location of the curvature functions (start and end points) is exactly identified along the track using the ideal geometry provided by the track preprocessor.

The curvature of the track can be experimentally approximated in the TGMS with the installed sensors. The curvature of the trajectory followed by the TGMS can be obtained as:

\begin{equation} \label{Eq8_1}
\rho _h^{\exp } \simeq \frac{{\hat \omega _z^{tgms}}}{V}
\end{equation}

Of course, this approximate measure is a \textit{noisy version} of the track horizontal curvature. However, experimental measures show that the overall shape of the curvature functions can be clearly obtained with this approximation.

\begin{figure}[htbp!]
	\centering
	\includegraphics[width=0.7\linewidth]{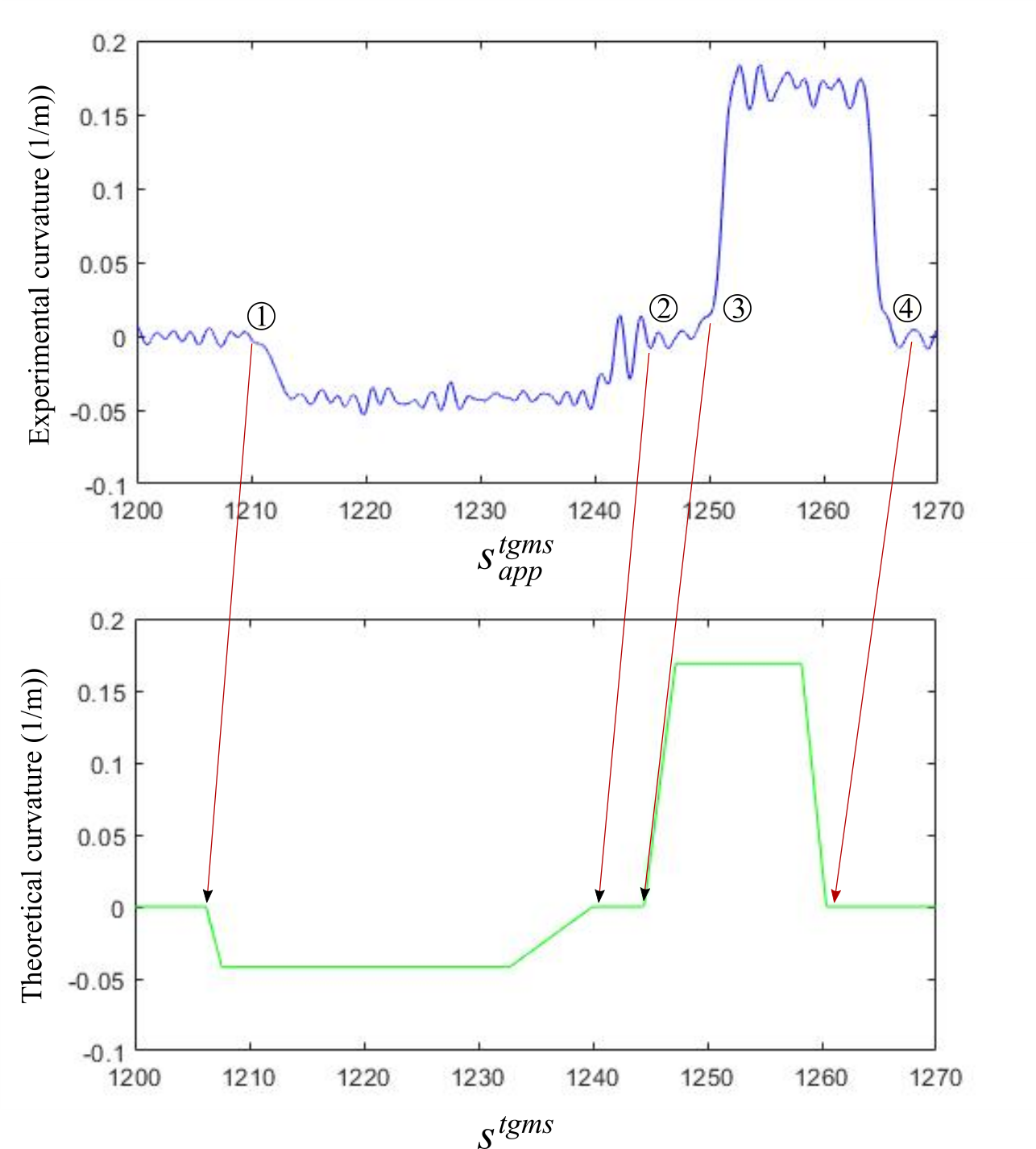}
	\caption{Odometry algorithm}
	\label{fig:fig8_2}
\end{figure}

\begin{figure}[htbp!]
	\centering
	\includegraphics[width=0.7\linewidth]{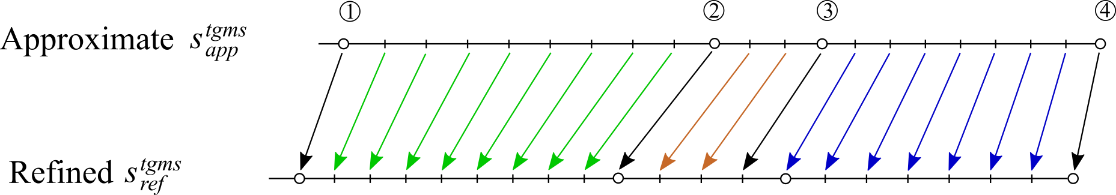}
	\caption{Correction of $^{tgms}$}
	\label{fig:fig8_3}
\end{figure}

The concept of the odometry algorithm, that is detailed in \cite{escalona2018odometry}, is to monitor the experimental curvature during the ride of the train using Eq. \ref{Eq8_1} and to store the data together with the approximate coordinate $s_{app}^{tgms}$ obtained with the help of the installed encoder. Plotting these data may look like the plot at the top of Fig. \ref{fig:fig8_2}. Using the track preprocessor, the ideal value of the curvature of the track $\rho _h^{ideal}$ in the area where the train is located, may look like the lower plot in Fig. \ref{fig:fig8_2}. As shown in the figure, this information can be used to correct the value of $s_{app}^{tgms}$ at points 1, 2, 3 and 4 located at the entry or exit of the curves. Measures of $s_{app}^{tgms}$ between these corrected points are also corrected using a linear mapping, as shown in Fig. \ref{fig:fig8_3}.

The problem is how to detect the entry and exit of the curves using the functions $\rho _h^{\exp }\left( {s_{app}^{tgms}} \right)$ and $\rho _h^{ideal}\left( {s^{tgms}} \right)$ . In fact, it is the exit of the curves what is detected first. Once the TGMS leaves a curve, the shape of the curvature function that the TGMS has ahead is known. Therefore, when the measured curvature $\rho _h^{\exp }\left( {s_{app}^{tgms}} \right)$ ”looks similar” to the expected curvature function $\rho _h^{ideal}\left( {s^{tgms}} \right)$, the exit of the curve has been reached. This similarity is computed by calculating at each instant the squared-error of the experimentally measured curvature and the expected curvature function, as follows:

\begin{equation} \label{Eq8_2}
e2\left( s \right) = {\int_{\bar s = s - \Delta s}^{\bar s = s} {\left[ {\rho _h^{\exp }\left( {\bar s} \right) - \rho _h^{ideal}\left( {\bar s - s + {s_{exit}}} \right)} \right]} ^2}\,d\bar s
\end{equation}

where $e2\left( s \right)$ is the squared error ($s$ substitute $s_{app}^{tgms}$  for simplicity in the formula), $\Delta s$ is the width of the expected curvature function and $s_{exit}$  is the location of the exit of the curve in the ideal geometry. For a better accuracy, the value of the squared error is normalized for each curvature function using the following factor:

\begin{equation} \label{Eq8_3}
I2 = {\int_{\bar s = 0}^{\bar s = \Delta s} {\left[ {\rho _h^{ideal}\left( {\bar s} \right)} \right]} ^2}\,d\bar s
\end{equation}

The normalization factors, that is different for each curvature function, is of course computed before the application of the method. The \textit{normalized squared-error} is given by: 

\begin{equation} \label{Eq8_4}
ne2\left( s \right) = \frac{{e2\left( s \right)}}{{I2}}
\end{equation}

\begin{figure}[htbp!]
	\centering
	\includegraphics[width=0.5\linewidth]{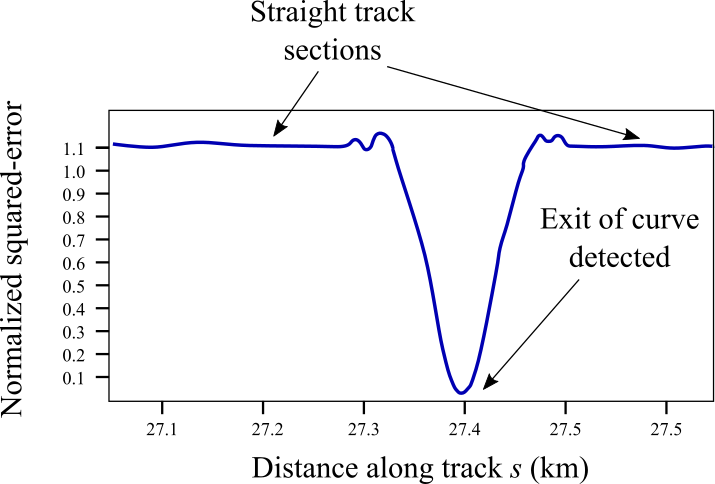}
	\caption{Normalized squared-error}
	\label{fig:fig8_4}
\end{figure}

Thanks to the normalization, the value of $ne2$ varies between approximately 1, in straight track sections, and 0 when there is a perfect matching between $\rho _h^{\exp }\left( {s_{app}^{tgms}} \right)$ and $\rho _h^{ideal}\left( {s^{tgms}} \right)$. A typical plot of the function is observed in Fig. \ref{fig:fig8_4}. This function uses to be smooth, such that detecting the local minimum that indicated the detection of the exit of the curve is a very easy task. Once the exit of the curve is detected, the expected curvature function is substituted by the next curve ahead along the track. It can be shown that this method is real-time capable. The main computational cost is the one associated with the calculation of the integral given in Eq. \ref{Eq8_2}.

\section{Sensor fusion algorithm to find TGMS to TF relative angles} \label{sec:sensFus}

The sensor fusion algorithm developed in this investigation is based on the Madgwick method \cite{madgwick2011sensFus}. The method that only uses the accelerometer and gyroscope data (IMU) without magnetometer data (no MARG) is used. In this method, the orientation of the sensor, that coincides with the orientation of the TGMS in the problem at hand, is obtained using the time-integration of the gyroscope signals and the direction of gravity obtained with the (capacitive) accelerometer signals. In this algorithm (and many other sensor fusion algorithms used to get orientation from an IMU) the following approximation for the accelerometer signals, given in Eq. \ref{Eq7_1} is used:

\begin{equation} \label{Eq9_1}
{{\bf{a}}^{imu}} = {\hat {\bf {\ddot  {R}}}^{tgms}} + {\left( {{{\bf{A}}^{tgms}}} \right)^T}{\left[ {\begin{array}{*{20}{c}}
0&0&g
\end{array}} \right]^T} \simeq {\left( {{{\bf{A}}^{tgms}}} \right)^T}{\left[ {\begin{array}{*{20}{c}}
0&0&g
\end{array}} \right]^T}
\end{equation}

this is, it is assumed that the component of the signal due to gravity is much larger that the component due to the sensor’s acceleration ($\left| {\hat {\bf {\ddot  {R}}}^{tgms}} \right| <  < g$). This approximation allows to find information about the rotation matrix ${{\bf{A}}^{tgms}}$ , that actually contains all information about the sensor’s spatial orientation, without knowing the value of ${\hat {\bf {\ddot  {R}}}^{tgms}}$. This approximation is not valid in our problem because accelerations due to the vehicle motion, like lateral accelerations in curves, can have significant values compared to $g$. 

The value of ${\hat {\bf {\ddot  {R}}}^{tgms}}$, whose expression is given in Eq. \ref{Eq7_1}, is unknown. However, an approximation to its value can be obtained as follows:

\begin{equation} \label{Eq9_2}
{\bar {\bf {\ddot  {R}}}^{tgms}} = {\bar {\bf {\ddot  {R}}}^{t}} + \ddot {\bf {\bar {r}}}^{tgms} + \left( \tilde{\bar{\alpha}}^t + \tilde{\bar{\omega}}^t\tilde{\bar{\omega}}^t \right){\bar{\bf {r}}^{tgms}} + 2{{\tilde {\bar {\omega}} }^t}\dot {\bf {\bar {r}}}^{tgms} \simeq {\bar {\bf {\ddot  {R}}}^{t}} = \left[ {\begin{array}{*{20}{c}}
{\dot V}\\
{{\rho _h}{V^2}}\\
{ - {\rho _v}{V^2}}
\end{array}} \right]
\end{equation}

This approximation is equivalent to assume that the acceleration of the TGMS is the one that a body moving along the ideal track with the same forward speed $V$ than the TGMS would experience. In other words, a particle moving along the thick line in Fig. \ref{fig:fig7_1} (“track centerline”) instead of the thin line (“TGMS trajectory”). The reader may think that this is a very rough approximation. However, results show that this approximate value is much more accurate than assuming $ {\bar {\bf {\ddot  {R}}}^{t}} \simeq {\bf{0}}$, as done in the original Madgwick method.

If the TGMS includes an inclinometer, its signal can be used to find more accurately the roll ${\varphi ^{tgms}}$  and pitch ${\theta ^{tgms}}$ angles. Inclinometers may not respond well at relatively high frequencies. A sensor fusion algorithm can be designed to calculate the low frequency component of these angles using the inclinometer signals and the high frequency component using the modified Madgwick algorithm described in this section. An example of such sensor function algorithm can be found in \cite{escalona2016verticalIrr}.

\section{Calibraton of the cameras} \label{sec:camCal}

\begin{figure}[htbp!]
	\centering
	\includegraphics[width=0.5\linewidth]{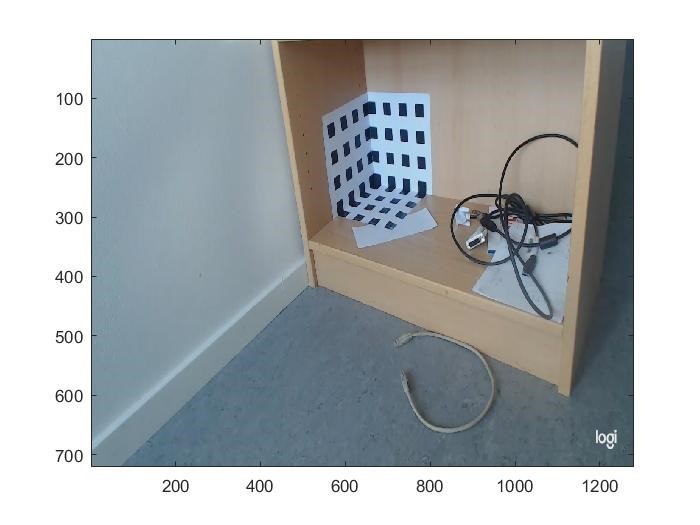}
	\caption{Calibration pattern used for camera calibration}
	\label{fig:fig10_1}
\end{figure}

As explained in Section \ref{sec:kinVis}, Eq. \ref{Eq4_1} is used to find the position vector of the points filmed in the camera. To that end, a set of intrinsic and extrinsic camera parameters have to be identified to build matrices ${{\bf{M}}^{int}}\,\,{\rm{and}}\,\,{{\bf{M}}^{ext}}$. The well-known Zhang’s method \cite{zhang1998camCal} can be used to this end. In \cite{escalona2020computerVision}, this method is adapted to the kinematic notation used in multibody dynamics. The calculation of the parameters based on the use of a trihedral pattern, as the one shown in Fig. \ref{fig:fig10_1}. is also explained in detail in \cite{escalona2020computerVision}.

The extrinsic parameters obtained when applying this method are the three components of the position vector of the camera with respect to the trihedral vertex (the corner of the shelf in the figure) and the three orientation parameters of the camera frame with respect to the trihedral frame (whose axis are aligned with the shelf edges). To that end, the inputs to the calibration process are:

\begin{enumerate}
    \item The 3D position vector of a set of points $P$ belonging to the pattern (at least five points in each of the three planes of the trihedral) with respect to the trihedral plane.
    \item	The position vector of the same set of points $P$ in the image in pixel coordinates.
\end{enumerate}

An optimization procedure can be followed to obtain ${{\bf{M}}^{int}}\,\,{\rm{and}}\,\,{{\bf{M}}^{ext}}$ as output. 

This procedure can be followed to calibrate the cameras of the TGMS.  However, the following points have to be considered:

\begin{enumerate}
    \item	In the TGMS, the position and orientation of the camera frame with respect to the trihedral frame are not of interest. What it is needed is the position vector ${\bf{\hat u}}_{cam}^{tgms}$ of the camera ($cam$ = $rcam$ or $lcam$) and the Euler angles ${\left[ {\begin{array}{*{20}{c}}
{{\varphi ^{tgms,cam}}}&{{\theta ^{tgms,cam}}}&{{\psi ^{tgms,cam}}}
\end{array}} \right]^T}$ of the camera frame with respect to the TGMS frame. These are the inputs of ${{\bf{M}}^{ext}}$ as shown in Eq. \ref{Eq4_2}.
\item	The position and orientation of the pattern with respect to the TGMS frame has to be precisely known in the calibration process. Otherwise, the calibration is useless.
\item	The parameters ${A^{las}},\,{B^{las}},\,{C^{las}}$ and ${D^{las}}$  that define the plane projected by the laser ($las$ = $rlas$ or $llas$) in the TGMS have to be identified in the calibration process.

\end{enumerate}

The calibration method proposed in this work is sketched in Fig. \ref{fig:fig10_2}. The calibration patterns have to be built on a structure that is rigidly connected to the TGMS (one for the left camera-laser equipment and one for the right camera-laser equipment). The inputs to the calibration process are the following:

\begin{figure}[htbp!]
	\centering
	\includegraphics[width=0.8\linewidth]{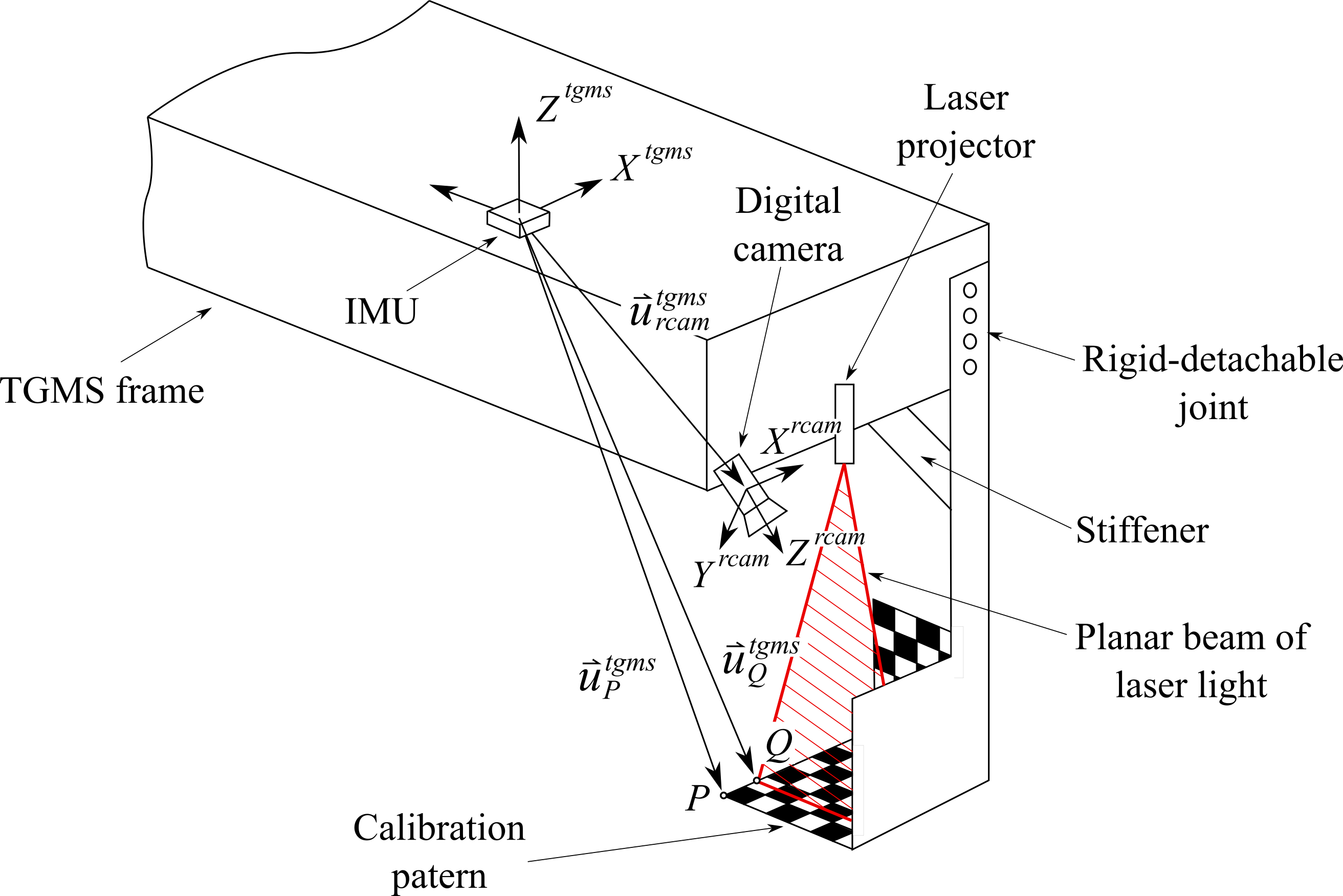}
	\caption{Installation of pattern in TGMS for calibration}
	\label{fig:fig10_2}
\end{figure}

\begin{enumerate}
    \item	The 3D position vector of a set of points $P$ (at least five points in each of the three planes of the trihedral) belonging to the pattern with respect to the TGMS frame.
\item	The 3D position vector of a set of points $Q$ (at least two points in two planes of the trihedral) belonging to the intersection of the laser beam with the pattern with respect to the TGMS frame.
\item	The position vector of the same set of points $P$ and $Q$ in the image in pixel coordinates.
\end{enumerate}

Using points $P$ as the input, just following the Zhang method explained in \cite{zhang1998camCal}, the extrinsic parameters ${\bf{\hat u}}_{cam}^{tgms}$ and ${\left[ {\begin{array}{*{20}{c}}
{{\varphi ^{tgms,cam}}}&{{\theta ^{tgms,cam}}}&{{\psi ^{tgms,cam}}}
\end{array}} \right]^T}$ (that in turn are used to calculate ${{\bf{A}}^{tgms,cam}}$) can be obtained. 

Following a simple optimization procedure based on the data of points $Q$, the laser plane parameters ${A^{las}},\,{B^{las}},\,{C^{las}}$ and ${D^{las}}$ can be easily identified.

For this calibration method to be accurate, the following conditions have to be followed:

\begin{enumerate}
    \item	The connection of the calibration pattern to the TGMS has to be machined in such a way their relative position is the same every time the calibration pattern is mounted. No clearances in the joint are allowed.
\item	The TGMS-calibration pattern has to form a rigid block when mounted. Structural stiffeners can be designed to that end, as shown in Fig. \ref{fig:fig10_2}.
\item	The positon vectors of points $P$ in the pattern with respect to the TGMS frame have to be determined accurately. To that end a \textit{coordinate measuring machine} (CMM) may be needed.
\item	The positon of points $Q$ in the TGMS frame can be obtained doing interpolation with the position vector of the “neighbor” points $P$.
\item	The position of the calibration patters with respect to the TGMS frame has to be as near as possible to the position of the rail cross-section with respect to the TGMS frame during the track geometry measurement. That why the cameras will be calibrated in useful area of the filmed frames. 

\end{enumerate}

\section{Summary of the measurement of track irregularities} \label{sec:summary}

The method explained in this document is now summarized.

\textbf{Input data:}
Each time-instant the following input data are needed:

\begin{enumerate}
    \item	Two digital-camera frames where the position vectors ${\bf{n}}_{P'}^{im}$ of points $P$ in the right rail and ${\bf{n}}_{Q'}^{im}$ of points $Q$ in the left rail can be detected.
\item	Three signals of the IMU accelerometer ${{\bf{a}}^{imu}}$ and three signals of the IMU gyroscope ${{\bf{\omega }}^{imu}}$. 
\item	An estimation of the position along the track $s_{app}^{tgms}$ obtained from the vehicle odometer, using a GNSS sensor or using an encoder in a wheel of the vehicle.
\item	Ideal geometry of the track and track preprocessor (computer program) to find the curvatures and slopes as a function of $s$.
\item	Optionally: inclinometer measure of ${\varphi ^{tgms}}$ and ${\theta ^{tgms}}$ angles.
\end{enumerate}

\textbf{Pre-process:}

Before starting the measurement of the track geometry, the cameras and laser projectors  have to be calibrated using the method explained in Section \ref{sec:camCal}. 

\textbf{Process:}

Each time instant, do:

\begin{enumerate}
    \item	Find ${\bf{\hat u}}_{Orrp}^{tgms}\,\,{\rm{and}}\,\,{\bf{\hat u}}_{Olrp}^{tgms}$ using the optimization method described in Section \ref{sec:detectXsec}.
\item	If $s_{app}^{tgms}$ is not accurate, calculate $s_{ref}^{tgms}$ using the odometry algorithm described in Section \ref{sec:odometry}.
\item	Calculate $V\,\,{\rm{and}}\,\dot V$ as the first and second numerical time-derivatives of $s_{ref}^{tgms}$, respectively.
\item	Use $s_{ref}^{tgms}$ to calculate the value of ${\rho _{tw}},\,\,{\rho _v},\,\,{\rho _h}\,\,{\rm{and}}\,\,{\rho '_h}$ of the track section.
\item	Use sensor fusion algorithm with corrected accelerations, as explained in Section \ref{sec:sensFus}, to find ${\left[ {\begin{array}{*{20}{c}}
{{\varphi ^{tgms}}}&{{\theta ^{tgms}}}&{{\psi ^{tgms}}}
\end{array}} \right]^T}$.
\item	\textbf{Calculate} relative irregularities, \textbf{gauge and cross-level}, using Eq. (6.5).
\item	Integrate Eq. \ref{Eq7_6} to find ${{\bf{\bar r}}^{tgms}}$.
\item	\textbf{Calculate} absolute irregularities, \textbf{alignment and vertical profile}, using Eq. \ref{Eq6_8}.
\end{enumerate}

\section{Final considerations} \label{sec:final}

In the introduction to this document it is mentioned that the twist irregularity of the track and the rail-head profile can be measured with this system. However, the process to find these measurements have not been mentioned in the body of the document. The reasons are:

\begin{enumerate}
    \item	The twist of the track is not really an independent irregularity. It is just a measure of the space-derivative of the cross-level. Therefore, if the cross-level is measured with sufficient space resolution, the twist can be easily obtained using numerical methods.
\item	The optimization process described in Section \ref{sec:detectXsec}, provides a comparison of the measured rail-head profiles and the ideal-unworn profile. Therefore, wear of the profiles at each filmed cross-section can be measured just overlapping both curves. It is true that the optimization process can be done using just a piece of the profile (the one that is observed in the frame). To get the complete rail-head profile, more than one camera per rail may be needed. However, wear of the rail head uses to occur just in the inner side (where flange-contact occurs). Therefore, a single camera watching the inner part of the rail-heads may be enough to measure the “interesting part” of the rail-head profile. 
\end{enumerate}


\end{document}